\documentclass[11pt,oneside,letterpaper]{article}
\usepackage{amssymb}
\usepackage{amsmath,bm}
\usepackage[dvips]{graphicx}
\usepackage{setspace}
\usepackage{fancyhdr}
\usepackage[dvipsnames]{xcolor}
\usepackage{ifpdf}
\usepackage{graphicx}
\usepackage{rotating}
\usepackage{comment}
\usepackage[margin=2.5cm]{geometry}
\usepackage[colorlinks=true, linkcolor  = blue,urlcolor=blue,citecolor=violet]{hyperref}
\usepackage{soul}
\usepackage{empheq}
\usepackage{bbold}
\usepackage{subfigure}
\usepackage{cite}

\def\J{{\cal{J}}}
\newcommand{\bi}{\begin{itemize}}
\newcommand{\ei}{\end{itemize}}
\newcommand{\bea}{\begin{eqnarray}}
\newcommand{\eea}{\end{eqnarray}}
\newcommand{\be}{\begin{equation}}
\newcommand{\ee}{\end{equation}}

\numberwithin{equation}{section}
\allowdisplaybreaks  

\begin{document}

\vspace*{2.5cm}
\begin{center}
{\LARGE \textsc{Constructing AdS$_2$ Flow Geometries}}
\end{center}

\vskip10mm
\begin{center}
{\small{Dionysios Anninos and Dami\'an A. Galante}}
\end{center}

\vskip 5mm
\begin{center}
{\small{\href{mailto:dionysios.anninos@kcl.ac.uk}{dionysios.anninos@kcl.ac.uk}, \quad \href{mailto:damian.galante@kcl.ac.uk}{damian.galante@kcl.ac.uk} }}
\end{center}

\vskip 5mm
\begin{center}
{\footnotesize{Department of Mathematics, King's College London, the Strand, London WC2R 2LS, UK}}
\end{center}
\vskip 5mm

\vspace{4mm}
 
\vspace*{0.6cm}

\vspace*{1.5cm}
\begin{abstract}
\noindent

We consider two-dimensional geometries flowing away from an asymptotically AdS$_2$ spacetime. Macroscopically, flow geometries and their thermodynamic properties are studied from the perspective of dilaton-gravity models. We present a precise map constructing the fixed background metric from the boundary two-point function of a nearly massless matter field. We analyse constraints on flow geometries, viewed as solutions of dimensionally reduced theories, stemming from energy conditions. Microscopically, we construct computationally tractable RG flows in SYK-type models at vanishing and non-vanishing temperature. For certain regimes of parameter space, the flow geometry holographically encoding the microscopic RG flow is argued to interpolate between two (near) AdS$_2$ spacetimes. The coupling between matter fields and the dilaton in the putative bulk is also discussed. We speculate on microscopic flows interpolating between an asymptotically AdS$_2$ spacetime and a portion of a dS$_2$ world.

\end{abstract}

\newpage
\setcounter{page}{1}
\pagenumbering{arabic}

\setcounter{tocdepth}{2}
\tableofcontents

\onehalfspacing

\section{Introduction \& summary}

In this paper we explore asymptotically AdS$_2$ geometries which are generally not isometric to a pure AdS$_2$ geometry in their interior. We refer to such spacetimes as flow geometries. Our main motivation in studying  flow geometries is to uncover how features living in the interior of the spacetime are encoded in observables anchored to the AdS$_2$ boundary, and to understand the microscopic interpretation of flow geometries with regard to their putative holographic dual. Our focus rests entirely on asymptotically AdS$_2$.
Although two-dimensional, AdS$_2$ is a geometry pertinent to the study of extremal and near-extremal black holes, including highly spinning ones. Moreover, working in two-dimensions opens the possibility of obtaining explicit expressions mapping boundary correlation functions to the complete non-linear bulk metric. Finally, we are interested in exploring these questions in a setting where the holographic dual is a quantum mechanical theory comprising a finite number of degrees of freedom residing at a sole point on a temporal worldline\footnote{One is reminded of the following line in Borges's ``The Aleph" (1945): {\it ``The Aleph?” I repeated. ``Yes, the only place on earth where all places are — seen from every angle, each standing clear, without any confusion or blending."}} rather than a quantum field theory endowed with spatial locality and a continuous infinity worth of degrees of freedom. In what follows we study flow geometries from both a macroscopic and microscopic perspective. 

The macroscopic picture is studied in the first three sections. In section \ref{grav1} we establish constraints on flow geometries stemming from energy conditions when viewed as the spherically symmetric sector of a four-dimensional spacetime. The main result of section  \ref{sec3} is a collection of formulas permitting the reconstruction of the non-linear bulk metric from the boundary two-point function of  nearly massless scalar fields. These formulas can be applied to a broad range of scenarios including time-dependent, thermal, and non-thermal spacetimes. In section \ref{sec_grav} we discuss dilaton-gravity models coupled to matter and review how to reconstruct the metric from thermodynamical considerations. 

In section \ref{sec5} we turn to the study of a particular microscopic model given by a two-flavoured SYK model subject to a relevant deformation. At large $N$ the deformed model at both vanishing and finite temperature admits an exact solution along the entire RG flow. Moreover, the undeformed model admits a nearly-marginal scalar operator $\mathcal{O}_\zeta$ in its low energy (near) conformal spectrum. From a bulk perspective $\mathcal{O}_\zeta$ corresponds to near massless scalar allowing us to connect with the results of section \ref{sec3}. The two-point function for $\mathcal{O}_\zeta$ is computed for both the undeformed and deformed model, at both vanishing and finite temperature. In section \ref{reconstructionsec} we connect the microscopic model to our macroscopic considerations. We reconstruct pieces of the putative bulk dual metric and dilaton-matter couplings. Finally, in section \ref{sec7} we discuss future directions and provide some speculative remarks. 

Below we provide a more detailed description of the main results.

\subsubsection*{\textbf{A. Macroscopic constraints $\&$ precise reconstruction} -- \ref{grav1}, \ref{sec3}, and \ref{sec_grav} }

Macroscopically, our interest lies in reconstructing a bulk feature -- such as the form of the metric or some other field deep in the interior of space -- from a set of boundary observables. Several approaches have been proposed to attack the macroscopic problem (see \cite{Hammersley:2006cp,Czech:2014ppa,Czech:2016xec,Engelhardt:2016wgb,Engelhardt:2016crc,Roy:2018ehv,Bao:2019bib,Hashimoto:2020mrx} for some recent work), though few focus specifically on asymptotically AdS$_2$. Often, these approaches rely on computing non-local objects such as minimal surfaces which are of interest due to the Ryu-Takayanagi relation \cite{Ryu:2006bv} to entanglement entropy in the boundary theory.

In the first part of the paper, we will study macroscopic properties of flow geometries and their reconstruction. We begin in section \ref{grav1} by establishing the constraints imposed on flow geometries when viewed as solutions of dimensionally reduced gravitational models coupled to matter satisfying certain energy conditions. Upon establishing their macroscopic viability, we proceed to build explicit maps from boundary Green functions to detailed features of bulk fields. As probes, we consider free matter fields propagating in a fixed two-dimensional background geometry $g_{\mu\nu}$ and dilaton field $\Phi$ in a given dilaton-gravity model. Up to boundary terms, the Euclidean action for the free matter scalar plus dilaton-gravity we consider is
\begin{equation}\label{dgm}
S_E = -\frac{1}{2\kappa}\int d^2x\sqrt{g} \left( \Phi R- \ell^{-2} U(\Phi) \right) - \frac{1}{2} \int d^2x \sqrt{g}\left( \mathcal{W}(\Phi)g^{\mu\nu} \partial_\mu \zeta \partial_\nu \zeta +  \mathcal{V}(\Phi) \zeta^2  \right)~.
\end{equation}
The form of $S_E$ entails a particular choice of frame in which the dilaton field has no derivatives. For a minimally coupled two-dimensional matter field $\zeta$, the functions $\mathcal{W}(\Phi)$ and $\mathcal{V}(\Phi)$ are constant. Generally they may depend on $\Phi$ through the functional form of $\mathcal{W}(\Phi)$ and $\mathcal{V}(\Phi)$. If the matter field stems from the dimensional-reduction of a higher-dimensional theory, $\mathcal{W}(\Phi)$ and $\mathcal{V}(\Phi)$ may be non-constant.

In full generality, we would like to understand how the precise form of the background fields $g_{\mu\nu}$ and $\Phi$, as well as the form of the functions $\mathcal{W}(\Phi)$ and $\mathcal{V}(\Phi)$ are encoded in the boundary observables of $\zeta$ and the thermodynamic features of the theory. We will make some steps toward accomplishing this goal. This is the subject of sections \ref{sec3} and \ref{sec_grav}. We find the following: 

\begin{itemize}

\item When $\mathcal{W}(\Phi)$ and $\mathcal{V}(\Phi)$ are constant, we construct a map between the boundary two-point function of $\zeta$ to the full non-linear bulk metric $g_{\mu\nu}$.  In order to do so in analytic form, we require that our fields must have a parametrically small mass in units of the AdS$_2$ length scale.  We can further construct this map for bulk metrics depending on both space and time. Explicit formulae mapping the boundary Green function to the bulk metric are given in (\ref{inverseLap}) and (\ref{2dspacetime}) respectively. We further provide concrete examples to test these.

\item In section \ref{wkb_analysis} we also consider cases where $\mathcal{W}(\Phi)$ and $\mathcal{V}(\Phi)$ are non-constant functions of $\Phi$. For $\mathcal{V}(\Phi) = 0$ we provide a systematic construction for $\mathcal{W}(\Phi)$ expanded near the AdS$_2$ boundary. If $\mathcal{W}(\Phi)$ is slowly varying, we also construct a map to linear order in the variation of $\mathcal{W}(\Phi)$, shown explicitly in (\ref{upsilonR}).

\item In section \ref{thermalrecon} we generalise the map to thermal geometries whose Euclidean boundary is given by an $S^1$.

\item In section \ref{sec_grav} we discuss the connection between the thermodynamic properties of the theory and the dilaton potential $U(\Phi)$. This is summarised in (\ref{thermo}). Furthermore, it is possible to find the two-dimensional metric $g_{ij}$ from $U(\Phi)$, using the map (\ref{schwDG}). Reconstructing the metric from thermodynamic considerations resonates with ideas in \cite{Jacobson:1995ab}.

\end{itemize}

\subsubsection*{\textbf{B. Microscopic RG flows $\&$ thermodynamics} -- \ref{sec5}}

Microscopically, the problem comes in two parts. The first is establishing that the candidate microscopic model, such as maximally supersymmetric Yang-Mills theory, indeed admits a holographic description. The full set of microscopic ingredients necessary for this to be the case is unclear. We do not address this problem. The second is computing the appropriate observables within a given state that allow for the desired reconstruction. Such observables may be simple to compute from a macroscopic starting point but rather forbidding from the microscopic perspective. So even if we have reason to believe our microscopic theory is holographic, it might be wise to reconstruct the bulk features from observables admitting computational control within the microscopic description! 

From the microscopic perspective, flow geometries correspond to states that break conformal invariance. To generate such states we can deform a microscopic model whose vacuum preserves conformality by a relevant deformation. Our motivation stems from the desire to understand the relation between a given relevant deformation of a microscopic model and the resulting macroscopic flow geometry. This may pave the way, for instance, toward designing interesting geometries in the interior of AdS$_2$ \cite{Anninos:2017hhn,Anninos:2018svg,Bena:2018bbd,Gross:2019ach,Gross:2019uxi}, or at least clarify the limitations thereof due to non-manifest constraints originating from consistency of a microscopic description. 

The microscopic model we analyse is an SYK-type model \cite{Sachdev:1992fk,kitaev_vid,Kitaev:2017awl,Maldacena:2016hyu} with two flavours \cite{Gross:2016kjj} deformed by a relevant deformation. The model is governed by a Hamiltonian of the following form \cite{Garcia-Garcia:2017bkg,Jiang:2019pam,Lunkin:2020tbq}
\begin{equation} \label{Hdeformed_intro}
H_{\text{def}} = H_{q} + s \, H_{q/2}~.
\end{equation}
Here, $H_{x}$ is an SYK Hamiltonian (\ref{genSYK}) with $x$ fermions of each flavour. The underformed model has $s=0$. It is constructed in such a way that it admits a near-marginal scalar operator $\mathcal{O}_\zeta$ given by (\ref{Ozeta}) in its low energy (near) conformal phase. 
At finite $s$, $H_{q}$ corresponds to a relevant deformation of the undeformed model significantly modifying the infrared behaviour of the original theory.

Remarkably, the RG flow of the large-$q$ model induced by the relevant deformation can be controlled analytically \cite{Jiang:2019pam} to leading order in the large $N$ limit. This remains the case for both vanishing and non-vanishing temperature. We establish regions of parameter space for which the flow is between two (near) conformal phases. At finite temperature, the specific heat exhibits linear dependence on the temperature around the two regimes corresponding to the (near) conformal phases. Once the RG flow has reached the first of the two conformal points, the theory is already in a strongly coupled phase.

\subsubsection*{\textbf{C. Holographic considerations} --  \ref{reconstructionsec}, and \ref{sec7}}

In \cite{Maldacena:2016upp,Jensen:2016pah,Gross:2017hcz} it has been argued that a certain low energy sector of SYK-type models is captured by dilaton-gravity theories with linear dilaton potential coupled to matter. Given the putative holographic nature\footnote{The SYK model has several features such as a large number of light states (including the $\mathcal{O}(N)$ fermions) which deviate from standard holographic models, and are suggestive of non-local bulk dynamics. Nevertheless, at least at sufficiently low temperatures/energies and for correlation functions with few insertions compared to $N$ a holographic picture seems pertinent. Identifying microscopic models of AdS$_2$ with bulk physics which is local  is an important outstanding problem. Fortunately, there is a wealth of AdS$_2$ solutions in string theory with local bulk dynamics which presumably admit a holographic description. It would be interesting to build a robust connection, along the lines of \cite{Anninos:2013nra,Anninos:2016szt} or otherwise \cite{Benini:2015eyy}, between the SYK literature to that of extreme horizons in string theory.} of the SYK model, we find a rare setting where a holographic RG flow \cite{deBoer:1999tgo} can be concretely explored from a microscopic starting point. Further to this, thermodynamic quantities such as the free energy are also calculable in our deformed microscopic model. Much like the low energy sector of the original SYK model, the low energy sector of the deformed model is in a strongly coupled phase. It seems reasonable to extend the hypothesis of \cite{Maldacena:2016upp,Jensen:2016pah,Gross:2017hcz} to the low energy sector of $H_{\text{def}}$. This is the subject of section \ref{reconstructionsec}, where we posit that the strongly coupled portion of the microscopic RG flow is macroscopically encoded in a deformation of the dilaton potential $U(\Phi)$ away from linearity. The form of $U(\Phi)$ is fixed by matching the thermodynamic quantities of the deformed microscopic theory to those of the putative dilaton-gravity model. The dilaton potential $U(\Phi)$ and corresponding bulk metric for the putative theory dual to the low energy sector of $H_{\text{def}}$ are described in section \ref{dilpotrec}. The geometric picture of the flow between two (near) conformal fixed points becomes a flow geometry interpolating between two (near) AdS$_2$ spacetimes. This is the two-dimensional counterpart of holographic RG flows studied, for example, in \cite{Freedman:1999gp}.

Additionally, the techniques developed throughout our macroscopic considerations in section \ref{sec3} are applied to the microscopic model (\ref{Hdeformed_intro}). This requires the presence of a weakly coupled bulk scalar field with parametrically small mass. Microscopically, a natural candidate is a `single trace' operator which is near marginal in the conformal phase. As already mentioned, the conformal phase of the underformed model contains such near marginal scalar operator $\mathcal{O}_\zeta$. Upon deforming the model, the large $N$ two-point function of $\mathcal{O}_\zeta$  (\ref{two_point_q}) can be computed along the full RG flow. The large $N$ limit further ensures that the interactions of the putative bulk matter fields are small. The coupling between the bulk matter fields and the dilaton-gravity fields is detailed in section \ref{bulkmattercoupling}. In the (near) conformal regions of the RG flow, the coupling is minimal in that $\mathcal{W}(\Phi)$ and $\mathcal{V}(\Phi)$ in (\ref{dgm}) are approximately constant. In the intermediate part of the RG flow, $\mathcal{W}(\Phi)$ acquires non-trivial functional dependence on $\Phi$ that we compute analytically close to each (near) conformal region. 

Finally, in section \ref{sec7} we discuss further research directions. We consider the analytically continuation of $s$ away from the real axis (further studied in appendix \ref{F_app}) and speculate on a potential relation to a bulk dual with a deep IR region characterised by a {\textit{positive}} cosmological constant \cite{Anninos:2017hhn,Anninos:2018svg}.

\section{Near extreme horizons, deformed.}\label{grav1}

We begin our discussion by considering a four-dimensional theory of gravity coupled to an electromagnetic field and a matter Lagrangian. The classical theory is described by the following Lorentzian action
\begin{equation}
S_L =  \frac{1}{16\pi G} \int d^4x \sqrt{-g} R  - \frac{1}{4} \int d^4 x\sqrt{-g} F_{\mu\nu}F^{\mu\nu} + S_{\text{matter}}~.
\end{equation}
There are several known solutions to the above theory. Our main interest will lie in those solutions describing black holes near the extreme limit of vanishing Hawking temperature. Such extremal black holes, when rotating, can already occur for the pure Einstein theory. For the sake of simplicity, and without suggesting the charged and rotating cases are equivalent \cite{Anninos:2017cnw,Anninos:2019oka,Castro:2018ffi,Moitra:2019bub,Chaturvedi:2018uov}, we consider the more symmetric, electrically charged solutions. In the absence of any matter fields turned aside from the electromagnetic field, the simplest solution takes the following form
\begin{equation}
ds^2 = -f(r) dt^2 + \frac{dr^2}{f(r)} + r^2 d\Omega_2^2~,  \quad\quad f(r) \equiv 1-\frac{2 M}{r} + \frac{Q^2}{r^2}~, \quad\quad A = \frac{Q}{r} \, dt~.  
\end{equation}
This is the Reissner-Nordstr\"om solution with ADM mass $M$ and electric charge $Q$, and we take $Q>0$ without loss of generality. If we analytically continue the geometry to Euclidean space, smoothness of the solution at the radial value $r_h$ satisfying $f(r_h)=0$ imposes that the Euclidean time coordinate is periodic. Absence of singularities also requires $M\ge Q$. If we take $\epsilon \equiv (M - Q)/Q$ and consider the limit $\epsilon \to 0$ at fixed $Q$, we see that $f(r)$ develops a double zero. This implies that a deep well is formed from any point in the interior of space down to $r=r_h\approx Q (1+\sqrt{2\epsilon}+\ldots)$. The size of this well scales as $-\log\epsilon$. The infinite throat allows us to expand the geometry near the point $r=r_h$ and obtain a new solution of the theory which describes only a small piece of the original space. In coordinates adapted to this near horizon region, to leading order in the small $\epsilon$-expansion we find the Bertotti-Robinson solution
\begin{equation}
\frac{ds^2}{Q^2} = - (\rho^2-1) d\tau^2 + \frac{d\rho^2}{(\rho^2-1)} + d\Omega_2^2~, \quad\quad\quad  A = - Q \, \rho \, d\tau~,
\end{equation}
where we have taken $r = Q(1 + \sqrt{2\epsilon}\rho)$, $t = Q \tau/\sqrt{2\epsilon}$ and $\rho\ge 1$. The above geometry is the product of AdS$_2$ and the round two-sphere.

It is of interest to understand whether there can be further structure within the near horizon geometry. When we are in the near extreme limit, this question can be sharpened to whether or not there exist geometries that asymptote to AdS$_2$, but are different as we enter the interior. A simple and somewhat natural starting point to consider this question is to establish any constraints imposed by the Einstein equations coupled to matter satisfying certain energy conditions.\footnote{A remarkable solution of the Einstein-Maxwell theory pertinent to our discussion was already discussed as early as 1947 by Majumdar \cite{PhysRev.72.390} and Papapetrou \cite{papapetrou}. 
Their solution, in the near horizon region, reads
\begin{equation}
ds^2 = -Q^4{V^{-2}(\bold{x})}{dt^2} +{V^2(\bold{x})}  {d\bold{x}^2}~, \quad V(\bold{x}) = \sum_{i=1}^n \frac{Q_i}{|\bold{x}- \bold{x}_i |}~, \quad\quad A(\bold{x}) = {V^{-1}(\bold{x})} dt~,
\end{equation}
where the $\bold{x}_i$ are fixed points on $\mathbb{R}^3$ and the total charge is $Q = \sum_i Q_i$. Expanding for $|\bold{x}| \gg |\bold{x}_i|$ we have
\begin{equation}
\frac{1}{|\bold{x}-\bold{x}_i|} = \sum_{l=1}^\infty |\bold{x}_i| \, P_l(\cos \theta_i) \times \frac{1}{|\bold{x}|^l}~,
\end{equation}
with $P_l(\cos \theta_i)$ the Legendre functions and $\theta_i$ the angle between $\bold{x}$ and $\bold{x}_i$. Then, we find the asymptotic behaviour 
\begin{equation}
\frac{ds^2}{Q^2} =   -r^2  \left(1 - \frac{\alpha(\boldsymbol{\theta})}{r}  + \ldots  \right) dt^2 +  \left(1 + \frac{\alpha(\boldsymbol{\theta})}{r} + \ldots \right)  \frac{dr^2}{r^2} + \left(1 + \frac{\alpha(\boldsymbol{\theta})}{r} + \ldots \right)d\Omega_2^2~,
\end{equation}
with $\alpha(\boldsymbol\theta) = 2 Q^{-2} \sum_i q_i P_1(\cos\theta_i)$. To leading order the above metric is the product of AdS$_2$ with the round two-sphere. The subleading corrections break the spherical symmetry as well as the $SL(2,\mathbb{R})$ symmetries of AdS$_2$. However, as $\bold{x}$ approaches any one of the $\bold{x}_i$ a new AdS$_2\times S^2$ throat forms. Given a total charge $Q$ and ADM mass $M$, there can be many solutions of this type corresponding to the large number of ways in which $Q$ can be partitioned.

Similar but richer solutions of nested AdS$_2$ throats arise as BPS solutions of four-dimensional $\mathcal{N}=2$ supergravity \cite{Denef:2000nb}. Microscopic considerations of such macroscopic configurations have been considered, for instance, in \cite{Denef:2002ru,Denef:2007vg,Anninos:2013nra,Bena:2012hf,Mirfendereski:2020rrk,anousnew}.
} Of these, we will focus on the null and weak energy conditions. 

\subsection{Constraints from the null energy condition}

If the deformation near the horizon preserves spherical symmetry we can consider the following setup. Take a family of static and spherically symmetric spacetimes
\begin{equation}\label{flowgeom}
ds^2 = - f(r) dt^2 + \frac{dr^2}{f(r)} + g(r)^2 d\Omega_2^2~,
\end{equation}
and further assume that the total matter stress tensor $T_{\mu\nu}$ obeys the null energy condition, namely that $T_{\mu\nu}v^\mu v^\nu \ge 0$ for all future pointing null vectors $v_\mu$. 
If the geometry (\ref{flowgeom}) is to obey the Einstein equation sourced by $T_{\mu\nu}$, one finds that $f(r)$ and $g(r)$ must obey certain inequalities. We assume that $f(r)$ and $g(r)$ are non-negative everywhere outside the position of the horizon $r=r_h$ where $f(r_h)= 0$ and $g(r_h) \neq 0$. Explicitly, 
\begin{equation}\label{ineq}
-f g g'' \ge 0~, \quad\quad 1- (g')^2 f +\frac{g^2 f''}{2} - f g g'' \ge 0~.
\end{equation}
The above inequalities relate the permitted behaviour in the flow of time, encoded in $f(r)$, and the flow of space, encoded in $g(r)$. They can be viewed as a spatial case of the Raychaudhuri equations. The weak energy condition would further require
\begin{equation}\label{weakineq}
1-g' \left(g f'+f g'\right)-2 f g g'' \ge 0~.
\end{equation}
The first inequality in (\ref{ineq}) is saturated for $g(r) = a + b r$, with $a$ and $b$ constant. If it is not saturated, the flow of the sphere size must be monotonically decelerating. The second inequality in (\ref{ineq}) simplifies if we define a new positive function $F(r) \equiv f(r)/g(r)^2$. It then reads
\begin{eqnarray}
(g^4 F')' \geq -2~, 
\end{eqnarray}
so the derivative of the function $g(r)^4 F'(r)$ is allowed to change sign as long as it does not become too negative. Note that this is possible because the transverse space is spherical. Planar or hyperbolic transverse spaces only allow for $(g^4 F')'\geq0$ as discussed in appendix \ref{nec}.

To have an asymptotically AdS$_2$ regime, we restrict $f(r) = (r-f_0)^2  + f_1 \log r + \ldots$ at large $r$.\footnote{The inclusion of a logarithmic term in the asymptotic expansion of the metric  may seem unusual. It is included for generality. In section \ref{sec5} we will study an SYK-type model subject to a relevant deformation, and in section \ref{reconstructionsec} we will argue that the geometry dual to the microscopic model has an asymptotic expansion (\ref{logmicro}) containing such a term.} 
We further permit $g(r) = g_0 + g_ 1 r + \ldots$ at large $r$, provided that $g_0 \gg g_1 r$ throughout the spacetime. Thus our spacetimes have a slowly varying sphere size, and might be more aptly characterised as asymptotically near AdS$_2$ geometries. The inequalities (\ref{ineq}) and (\ref{weakineq}) indeed admit such functions.
More general considerations of asymptotically AdS$_2$ spacetimes can be found, for example, in \cite{Cvetic:2016eiv,Castro:2008ms,Galloway:2018dak,Tod:2018qmp}.
\newline\newline
\textbf{AdS$_2$ to AdS$_2$.} As a simple example, we take
\begin{equation}
f(r) = \left( \frac{1}{r} + \frac{1}{r-1}\right)^{-2}~, \quad\quad\quad g(r) = 1 + \frac{r}{r_0}~,
\end{equation}
with $r\ge 1$ and $r_0 > 1$. If we further consider $r_0 \gg 1$, we can consider the limit $r_0 \gg r\gg 1$ where the geometry is approximated by
\begin{equation}
ds^2 \approx - \frac{r^2}{4}\left(1- \frac{1}{r} + \ldots \right) dt^2 + \frac{4}{r^2} \left(1+\frac{1}{r}  + \ldots \right) dr^2 + \left(1 + \frac{2r}{r_0}  + \ldots\right) d\Omega_2^2~. 
\end{equation}
On the other hand, near $r=1$ we have
\begin{equation}
ds^2 \approx - {z^2}\left(1- 2z  + \ldots \right) dt^2 +\frac{1}{z^2}\left(1+2 z  + \ldots\right) dz^2 + \left(\left(1+ \frac{1}{r_0}\right)^2 + \ldots \right) d\Omega_2^2~,
\end{equation} 
where $z = r-1$. So the geometry interpolates between two AdS$_2$ like regimes. Taking $r_0\to\infty$ we see that these interpolating solutions can even have a constant two-sphere size along the radial direction. In addition to spacetimes interpolating between two AdS$_2$ like regimes, one can find even more exotic behaviour compatible with (\ref{ineq}). 
An instance of this is a geometry interpolating from an asymptotically AdS$_2 \times S^2$ boundary to a portion of dS$_2 \times S^2$ in the interior. 
\newline\newline
To conclude, the AdS$_2$ region near the horizon of an extreme black hole may admit deformations compatible with the null energy condition encoding a wealth of features in its interior. In section \ref{sec5} we will consider such deformations from a more microscopic perspective. In light of this, it is useful to consider how these deformations appear from the perspective of observables anchored at the AdS$_2$ boundary.  

\section{Deformed near extreme horizons, marginally explored.} \label{sec3}

In this section we derive explicit relations between boundary correlators and the bulk metric. We restrict our attention to the class of spherically symmetric metrics
\begin{equation}\label{4dconf}
ds^2 = e^{2\gamma(t,z)} \left( dt^2 + dz^2 \right) + e^{2\upsilon(t,z)} d\Omega_2^2~.
\end{equation}
We have chosen a conformal frame for the two-dimensional part of the metric, and we take $z \in \mathbb{R}^+$ and $t\in\mathbb{R}$. We work in Euclidean signature. We take $z=0$ to be the AdS$_2$ boundary while $z\to\infty$ is the deep interior of space. Our main results are expressions (\ref{inverseLap}), (\ref{2dspacetime}) where the bulk metric is reconstructed for arbitrary $\gamma(t,z)$ with constant $\upsilon(t,z)$. In (\ref{upsilonR}) we further reconstruct a slowly varying $\upsilon(t,z)$.

From now on, we will work in units where the radius of AdS$_2$ is set to unity unless otherwise specified.  The pure AdS$_2$ geometry is given by $e^{-2\gamma(t,z)} = z^2$. As our probe, we consider a minimally coupled massive scalar field, $\zeta$, with quadratic action given by
\begin{equation}
S_{\text{probe}} = \frac{1}{2} \int d^4 x \sqrt{g}  \left( g^{\mu\nu} \partial_\mu \zeta \partial_\nu \zeta + m^2 \zeta^2  \right)~.
\end{equation}
 We begin by considering the static case, with $\gamma(t,z) = \gamma(z)$ and $\upsilon(t,z) = \upsilon(z)$. The sector of the matter theory that is independent of the two-sphere is described by the two-dimensional theory
\begin{equation}\label{probe4d}
S^{(2\text{d})}_{\text{probe}} = \frac{1}{2} \int dt dz e^{2\upsilon(z)} \left(  \partial_t \zeta \partial_t \zeta + \partial_z \zeta \partial_z \zeta   + e^{2\gamma(z)} m^2 \zeta^2  \right)~,
\end{equation}
where now $\zeta = \zeta(t,z)$ is a two-dimensional field. Had we considered a non-minimal coupling to the background Ricci scalar or electromagnetic field strength, the form of the two-dimensional action governing the spherically symmetric sector would remain unchanged but the interpretation of $\gamma(z)$ and $\upsilon(z)$ would differ. 

The wave-equation of the spherically symmetric sector is given by
\begin{equation}\label{zetaKG}
\left(- e^{-2\upsilon(z)} \partial_z e^{2\upsilon(z)}\partial_z + \omega^2 + e^{2\gamma(z)}m^2 \right)\zeta_\omega(z) = 0~,
\end{equation}
where we have performed a Fourier expansion in $t$ with conventions 
\begin{equation}
f(t) = \int_{\mathbb{R}} \frac{d\omega}{2\pi} e^{-i\omega t} f_\omega~.
\end{equation}

The observable we will be interested in will be the boundary correlator of $\zeta$. This can be calculated by evaluating the on-shell action $S^{(2\text{d})}_{\text{probe}}[\xi_\omega]$ of a solution with boundary behaviour $\xi_\omega$ at a small value $z = z_c$. One readily obtains
\begin{equation}\label{probeS}
S^{(2\text{d})}_{\text{probe}}[\xi_\omega] = - \frac{1}{2} \int_{\mathbb{R}} \frac{d\omega}{2\pi} \xi_\omega \, e^{2\upsilon(z_c)}  \partial_z \zeta_{-\omega}(z) |_{z=z_c}~,
\end{equation} 
from which the boundary two-point function can be computed by
\begin{equation} \label{deltaZ}
G(\omega,\omega') = \left. \frac{\delta^2}{\delta \xi_\omega \, \delta \xi_{\omega'} }  \log Z[\xi_\omega] \right|_{\xi_{\omega}=0} \,,
\end{equation}
where $Z[\xi_\omega] = e^{-S^{(2\text{d})}_{\text{probe}}[\xi_\omega]}$. Since we have time-translation invariance, in the static case $G(\omega,\omega') = G(\omega)\delta(\omega+\omega')$.  We do not know how to solve the wave-equation (\ref{zetaKG}) analytically for arbitrary $\gamma(z)$ and $\upsilon(z)$. To make progress, we have to make further simplifying assumptions. In what follows we discuss various such simplifications. 

\subsection{Small mass limit with constant two-sphere size}

 As the simplest possible scenario, one might imagine that the field is massless and furthermore that $\upsilon(z) = \upsilon_0$ is constant. This scenario is too simple however. A minimally coupled massless scalar in two-dimensions is conformal and does not perceive the conformal factor of the geometry, so it cannot be a good probe for it. 
 
 As a second simplest scenario, we can keep $\upsilon(z) = \upsilon_0$ constant and assume that $m$ is small in units of the AdS$_2$ length scale. We would like to setup a perturbative analysis. To leading order in this perturbative approach, we simply solve the massless equation. Choosing the solution that is regular in the interior yields
\begin{equation}
\zeta^{(0)}_\omega(z) = \xi_\omega \, e^{-|\omega| (z-z_c)}~.
\end{equation}
From the above solution we can derive the Green's function obeying Dirichlet boundary conditions at $z=z_c$
\begin{equation}
G(\omega;z,y) = \Theta(y-z) \frac{\sinh  \omega (z-z_c)}{\omega} \, e^{-|\omega|(y-z_c)} + \Theta(z-y) \frac{\sinh \omega (y-z_c)}{\omega} \, e^{-|\omega|(z-z_c)}~.
\end{equation}
Treating the mass term in the wave equation as a small perturbation, we can find the corrected solution. One has
\begin{equation}
\zeta_\omega(z) = \xi_\omega \left( e^{-|\omega| (z-z_c)} - m^2 \, \int^\infty_{z_c} dy  \, G(\omega;z,y) e^{2\gamma(y)}e^{-|\omega| (y-z_c)} + \ldots \right) \,.
\end{equation}
Explicitly,
\begin{multline}
\zeta_\omega(z) = \xi_\omega  e^{-|\omega| (z-z_c)} - \xi_\omega m^2 \frac{\sinh  \omega (z-z_c)}{\omega} \,  \int^\infty_{z} dy  \, e^{2\gamma(y)}e^{-2|\omega| (y-z_c)} \\ 
  - \xi_\omega m^2 \frac{e^{-|\omega |(z-z_c)}}{\omega} \,  \int^z_{z_c} dy \, e^{2\gamma(y)} e^{-|\omega| (y-z_c)} \sinh \omega(y-z_c) +
\ldots 
\end{multline}
We can expand the above solution near $z=z_c$ to find
\begin{equation} \label{bdy_zeta}
\zeta_\omega(z) \approx \xi_\omega \left( 1 - \left( |\omega| + m^2  \int_{z_c}^{\infty } e^{2 \gamma (y)}e^{-2 \left| \omega \right|  \left(y-z_c\right)} \, dy \right) \left(z-z_c\right)   +O\left( (z-z_c)^2 \right)+ \cdots \right) \,.
\end{equation}
By plugging (\ref{bdy_zeta}) into (\ref{deltaZ}) we directly read off the boundary two-point function
\begin{equation}\label{greenbdy}
G(\omega) = -e^{2\upsilon_0} \left( |\omega| + m^2  \int^\infty_{z_c} dy  \, e^{2\gamma(y)} \,e^{-2|\omega| (y-z_c)} + \ldots \right) \,.
\end{equation}
We note that the above expression, though perturbative in $m^2$, is non-linear in the conformal factor $\gamma(z)$.

It is straightforward to generalise the above discussion to the time-dependent case where now $\gamma = \gamma(t,z)$. In this case,  (\ref{greenbdy}) must be replaced with
\begin{equation}\label{greenbdyii}
{G}(\omega_1,\omega_2) = -e^{2\upsilon_0}|\omega_1| \delta(\omega_1-\omega_2) - e^{2\upsilon_0} m^2   \int^\infty_{z_c} dy  \,\Gamma_{\omega_2-\omega_1}(y) \,e^{-(|\omega_1|+|\omega_2|) (y-z_c)} + \ldots~,
\end{equation} 
where $\Gamma_l(z)$ is the Fourier transform of $e^{2\gamma(t,z)}$. Note that now the boundary two-point function depends on two frequencies.
\newline\newline
{\textbf{Example}.} As a simple example, we can consider the pure AdS$_2$ geometry for which $e^{2\gamma(z)} = z^{-2}$. We find the non-local part of the boundary two-point function
\begin{equation}\label{pureadsG}
\tilde{G}(\omega) =  - |\omega| - 2 m^2 |\omega| \log |\omega| z_c + \ldots  + \text{local}
\end{equation}
where we have defined $\tilde{G} (\omega) \equiv G(\omega) e^{-2\upsilon_0}$ for convenience. The local terms are non-negative powers of $\omega$. It is readily checked that the above is indeed the small mass expansion of the known expression at finite mass.\footnote{ The two-point function for arbitrary mass \cite{Freedman:1998tz} is  given by  $G(\omega) = 2^{1-2 \Delta }\frac{ \Gamma \left({1}/{2}-\Delta \right)}{\Gamma \left(\Delta -{1}/{2}\right)} \left| \omega \right| ^{2 \Delta -1}$ with $\Delta = 1/2 + \sqrt{m^2+1/4}$.} A slightly more involved example is shown in appendix \ref{app_centaur}, where we compute the two-point function for the centaur geometry \cite{Anninos:2017hhn,Anninos:2018svg}.

\subsection{Mapping from boundary correlator to bulk metric}

We could also ask the reverse question. Given $G(\omega,\omega')$, is it possible to efficiently reconstruct $\gamma(t,z)$? We begin by considering a constant two-sphere size.

Consider first the static case. In the small mass limit, this question amounts to whether we can invert (\ref{greenbdy}). This is very similar to performing an inverse Laplace transform. Care must be taken with the $z_c\to0$ limit however, since expressions may diverge. These divergences can be softened by taking $\omega$ derivatives. Let us further assume, for concreteness, that $\gamma(z)$ admits an expansion near $z=0$ of the type
\begin{equation}
e^{2\gamma(z)} = {z^{-2}} \left(1 +  \ldots \right)~, 
\end{equation}
where the dots can be any function of $z$ that goes to zero as $z\to 0$.
It then follows that we can take $z_c\to0$ for $\partial^2_\omega G(\omega)$ without encountering any divergences. We find
\begin{equation}\label{inverseLap}
\boxed{e^{2\gamma(z)} = - \frac{1}{8\pi i \, m^2} \times \frac{1}{z^2} \, \int_{-i\infty+ \delta}^{+i\infty +  \delta} d\omega \, e^{2\omega z} \, \partial^2_\omega \tilde{G}(\omega)}
\end{equation}
where $\partial_\omega^2 \tilde{G}(\omega)$ is understood to be analytically continued from the positive real $\omega$-axis to the complex $\omega$-plane. Here $\delta \in \mathbb{R}^+$ is larger than the real part of any poles in $\partial_\omega^2 \tilde{G}(\omega)$. The analytic structure of $\tilde{G}(\omega)$ is encoded in the analytic structure of $\gamma(z)$ through (\ref{greenbdy}). It is clear from (\ref{inverseLap}) that the large $\omega$ expansion of $G(\omega)$ can be used to systematically produce a small $z$ expansion for $\gamma(z)$. This follows from the fact that the inverse Laplace transform of $\omega^{-n}$ is
\begin{equation} \label{laplace_omega}
 \frac{1}{2\pi i}\int_{-i\infty+ \delta}^{+i\infty +  \delta} d\omega \, e^{2\omega z} \, \omega^{-n} = \frac{2^n}{\Gamma(n)}\times z^{n-1}~.
\end{equation}
The small $\omega$ expansion of $G(\omega)$, on the other hand, encodes the large $z$ behaviour of $\gamma(z)$. However, we do not know of a systematic procedure to relate the small $\omega$ expansion to the large $z$ behaviour of $\gamma(z)$. It would be interesting to explore this.

It is also possible to invert the time-dependent formula (\ref{greenbdyii}). For that, it is convenient to define $\omega_+ \equiv \omega_2 + \omega_1$ and $\omega_- \equiv \omega_2 - \omega_1$. We can then reconstruct the Fourier coefficients $\Gamma_{\omega_-}(z)$ of $e^{2\gamma(t,z)}$ in a similar way:
\begin{equation}\label{2dspacetime}
\boxed{\Gamma_{\omega_-} (z) = - \frac{1}{8\pi i \, m^2} \times \frac{1}{z^2} \int^{i \infty +\delta}_{-i \infty +\delta} d\omega_+ e^{2 \omega_+ z} \partial^2_{\omega_+} \tilde{G}(\omega_+,\omega_-) \,}
\end{equation}
Again, $\delta$ is larger than the real part of any poles in $\partial_{\omega_+}^2\tilde{G}(\omega_+,\omega_-)$ in the complex $\omega_+$-plane at fixed and real $\omega_-$. 
\newline\newline
\textbf{Example.} As a slightly non-trivial analytically tractable application of the above formula, we take the case $e^{-2\gamma(z)} = \tanh^2 z$. For this case, we have
\begin{equation}
-\tilde{G}(\omega) = \left(1 + 2m^2 \log \left(  2 e^{\gamma_E} z_c \right) \right) |\omega| +  \frac{m^2}{2|\omega|} + 2 m^2 |\omega| \, \psi^{(0)}(|\omega| )  +  \, \text{local}~,
\end{equation}
where $\psi^{(0)}(\omega)$ is the digamma function and $\gamma_E$, the Euler's constant. The second derivative is found to be
\begin{equation}
\partial_\omega^2 \tilde{G}(\omega) = -m^2 \left( {|\omega|^{-3}}+4 \psi ^{(1)}(|\omega| )+2 |\omega|  \psi^{(2)}(|\omega| )\right)~.
\end{equation}
Here $\psi^{(n)}(\omega)$ is the $n$-th derivative of the digamma function. Computing the inverse Laplace transform of the above equation we recover $e^{-2\gamma(z)} = \tanh^2 z$. At large $\omega$ we have
\begin{equation}
\tilde{G}(\omega) = - |\omega| - 2m^2 |\omega| \log |\omega|z_c - \frac{m^2}{3}  \frac{1}{|\omega|} + \ldots  + \text{local}
\end{equation}
The first two terms are the same as in the pure AdS$_2$ case (\ref{pureadsG}). The third term indicates that the geometry is no longer pure AdS$_2$ and encodes information about the small $z$ behaviour of $\gamma(z)$. In fact, using (\ref{laplace_omega}), it is straightforward to see that the $\sim |\omega|^{-1}$ term corresponds to the constant term in the expansion of the metric,
\begin{equation}
e^{2\gamma(z)} = \frac{1}{\tanh^{2} z} = \frac{1}{z^{2}} + \frac{2}{3} + O(z^2).
\end{equation}

\subsubsection*{Alternate boundary conditions}

When considering light fields in AdS, we can also consider the possibility of alternate boundary conditions. We would like an expression for the generating function of correlation functions in the theory where the fixed boundary data correspond to the profile of the fast-falling mode. One way to achieve this \cite{Klebanov:1999tb, Mueck:2002gm, Gubser:2002vv} is to perform the following Legendre transform of the Euclidean partition function 
\begin{equation}
\log Z_{\text{alt}}[\eta(t)] =  - \int_{\mathbb{R}} dt \eta(t) \xi(t) +  \log Z_{\text{Dir}}[\xi(t)]~,
\end{equation}
where we relate $\xi(t)$ to $\eta(t)$ through
\begin{equation}
\eta(t) = - \frac{\delta}{\delta \xi(t)} \log Z_{\text{Dir}}[\xi(t)]~, \quad\quad \xi(t) =  \frac{\delta}{\delta \eta(t)} \log Z_{\text{alt}}[\eta(t)]~.
\end{equation}
For a scalar in Euclidean AdS$_2$ of mass $m$, the boundary Green's function for the alternate boundary conditions is related in a simple way to the Dirichlet case
\begin{equation}\label{Galt}
G_{\text{alt}}(\omega) = -\frac{(2\Delta-1)^2}{G(\omega)}~, \quad\quad \Delta = \frac{1}{2}+\sqrt{m^2 + \frac{1}{4}}~.
\end{equation}
Thus, given the boundary Green's function for the alternate boundary condition, we can use (\ref{Galt}) to map the problem to the Dirichlet case and  apply (\ref{inverseLap}) to reconstruct the two-dimensional Weyl factor $e^{2\gamma(z)}$.

\subsection{Massless field with varying two-sphere size} \label{wkb_analysis}

The next-to-simplest case is given by no longer assuming that $\upsilon(z)$ is constant, while taking $m$ to be massless. We would like to solve the following wave-equation
\begin{equation}\label{schro}
\left(-\partial^2_u  +  e^{4\upsilon(u)}  \omega^2 \right) \zeta^{(0)}_\omega(u) = 0~,
\end{equation}
where we have introduced the follwoing $u$-coordinate
\begin{equation}
u = \int_0^z d\tilde{z}e^{-2\upsilon(\tilde{z})}~,
\end{equation}
in order to put the wave-equation in Schr\"odinger form. The coordinate $u$ grows monotonically with $z$ and $u=0$ at $z=0$. We can study (\ref{schro}) in a WKB limit where $\omega$ becomes large, see appendix \ref{wkb_app}. To leading order
\begin{equation}
\zeta_\omega^{(0)}(u) = \xi_\omega  e^{-(|\omega| z(u)+(\upsilon(u)-\upsilon_0) + \ldots)} \,,
\end{equation}
where $\upsilon_0 \equiv \upsilon(0)$.
Using the recursion relation for higher order terms in the WKB expansion we can systematically construct higher order corrections (see for instance \cite{PhysRevD.16.1740}). Using the WKB expansion and the on-shell action (\ref{probeS}) we can obtain a large $\omega$ expansion of the boundary two-point function. We find
\begin{equation}
G(\omega) =   -e^{2\upsilon_0} |\omega|  - \upsilon'(0) + O (|\omega|)^{-1}  ~.
\end{equation} 
Further progress can be made if we assume the sphere is near-constant as a function of $u$, which we parameterise as
\begin{equation}
\upsilon(u) = \upsilon_0 + \varepsilon \delta\upsilon(u)~,
\end{equation}
with $\varepsilon$ small. This can be viewed as a small correction away from exact extremality. To leading order in a small $\varepsilon$ expansion, (\ref{schro}) can be solved in a similar manner to the previous case. The resulting expression for the Green's function is
\begin{equation}
\tilde{G}(\omega) = - |\omega| - 4 \, \varepsilon \, e^{2\upsilon_0}  \, \omega^2   \int^\infty_{u_c} dy  \, \delta \upsilon(y) \, e^{-2|\omega| e^{2\upsilon_0} (y-u_c)} + \ldots \,.
\end{equation}
This formula can be inverted, but as in the previous case, we need to take care of potential divergences when $u\to0$. Assuming $\delta \upsilon = u^{-1} (1+ \cdots)$,
the inverse expression is
\begin{equation}\label{upsilonR}
\delta \upsilon(u) =  \frac{e^{-2\upsilon_0}}{8\pi i \, \varepsilon}  \frac{1}{u} \, \int_{-i\infty+ \delta}^{+i\infty +  \delta} d\omega \, \exp \left(2 e^{2\upsilon_0} \omega u \right) \, \partial_\omega \left( \frac{\tilde{G}(\omega)}{\omega^2} + \frac{1}{\omega} \right) \,.
\end{equation}
As before, $\delta \in \mathbb{R}^+$ needs to be greater than the real part of any poles in the integrand. Along the lines of (\ref{greenbdyii}), it is also possible to find a generalised expression for the time-dependent case where $\delta \upsilon = \delta \upsilon(t,u)$. \newline\newline
\textbf{Example.} As a simple example consider the case $\delta\upsilon(u) =  a u^{-1}$. This gives rise to
\begin{equation}
\tilde{G}(\omega) = - |\omega| + 4 \, \varepsilon \, a \, e^{2\upsilon_0} \omega ^2  \log |\omega|  + \text{local} +  \ldots \,,
\end{equation}
where we have kept the leading non-local term in the small the $u_c$-expansion.

\subsection{Extension to thermal spacetimes}\label{thermalrecon}

So far we have only considered the case for which the Euclidean AdS$_2$ boundary is the real line, parameterised by $t\in\mathbb{R}$. It is natural to also consider an $S^1$ boundary, parameterised by $t \sim t + 2\pi$. Euclidean AdS$_2$ with an $S^1$ boundary is the Poincar\'e disk with metric $e^{-2\gamma(z)} = \sinh^2 z$. Equation (\ref{greenbdy}) for the perturbative correction of the boundary Green's function remains unchanged except that the frequencies are now restricted to $\omega_n = 2\pi n$ with $n \in \mathbb{Z}$. In order to be able to reverse engineer the metric and use (\ref{inverseLap}), we must be able to extend $G(\omega_n)$ to the full real line. By Carlson's theorem, this can be done in a unique way provided that the analytic extension does not grow too rapidly at infinity. Similar considerations hold for (\ref{upsilonR}).

As a simple example we take $e^{-2\gamma(z)} = \sinh^2 z$ in the small mass limit with constant two-sphere size. Using (\ref{greenbdy}), we find 
\begin{equation}\label{thermaleq}
\tilde{G}(\omega_n) = - |\omega_n| - 2 m^2  |\omega_n|  \left( \psi ^{(0)}(|\omega_n| ) + \log \left(2 e^{\gamma_E} z_c\right)\right) + \ldots
\end{equation}
We can compare this to the small mass expansion of the exact result \cite{Sachdev:2015efa}
\begin{equation}
\tilde{G}_{\text{exact}}(\omega_n) = \frac{2^{1-2 \Delta } \Gamma \left(\frac{1}{2}-\Delta \right)}{\Gamma \left(\Delta -\frac{1}{2}\right)} \frac{\Gamma(\Delta+|\omega_n|)}{\Gamma(1-\Delta+|\omega_n|)}~,
\end{equation}
and find agreement. The function (\ref{thermaleq}) can be uniquely continued to a function $\tilde{G}(\omega)$, across $\omega\in\mathbb{R}$ obeying the conditions of Carlson's theorem. Consequently, we can use (\ref{inverseLap}) with $\tilde{G}(\omega)$ to retrieve $e^{-2\gamma(z)} = \sinh^2 z$.

\section{Two-dimensional dilaton-gravity, plus matter.} \label{sec_grav}

In this section we discuss dilaton-gravity models in two-dimensions coupled to matter fields (for some references see \cite{Cavaglia:1998xj, Grumiller:2007ju, Anninos:2017hhn, Witten:2020ert}). These might be viewed as the dimensional reduction of some higher-dimensional theory. However, the models can also be viewed as simple stand-alone models of two-dimensional gravity. Our motivation for focusing on such models comes from recent progress on a class of microscopic quantum mechanical models that bear a holographic relation to dilaton-gravity theories.

\subsection{Euclidean action and thermodynamics}

The Euclidean action governing the dilaton-gravity sector of the theory on a manifold $\mathcal{M}$ with disk topology is given by
\begin{equation}\label{DG}
S_E = - \frac{1}{2\kappa} \int_{\mathcal{M}} d^2x \sqrt{g}\left( \Phi R - {\ell^{-2}}{U(\Phi)}  \right) - \frac{1}{\kappa} \int_{\partial\mathcal{M}} du \sqrt{h} \Phi_b K~,
\end{equation}
where $K$ is the extrinsic curvature and $h$ is the induced metric at $\partial\mathcal{M}$ and $\Phi_b$ is the value of $\Phi$ at $\partial\mathcal{M}$. We have chosen a frame in which the action contains no $\Phi$ derivatives. In addition, we can always add a purely topological term
\begin{equation}
S_{\text{top}} = -\frac{\Phi_0}{2\kappa} \int d^2x \sqrt{g} R - \frac{\Phi_0}{\kappa}\int_{\partial\mathcal{M}} du \sqrt{h} K~,
\end{equation}
with $\Phi_0$ being constant. It can be shown \cite{Cavaglia:1998xj} that the theory (\ref{DG}) admits the following static solution
\begin{equation}\label{schwDG}
\frac{ds^2}{\ell^2} = f(r) d\tau^2  + \frac{dr^2}{f(r)}~, \quad\quad f(r) = \frac{1}{\tilde{\Phi}} \int_{r_h}^r d\tilde{r} \, U(\Phi(\tilde{r}))~, \quad\quad  \Phi(r) = \tilde{\Phi} \, r~.	
\end{equation}
The origin of the disk lies at $r=r_h$, which is the Euclidean horizon. The function $f(r)$ must be everywhere positive. Moreover, at least within a finite neighbourhood, all solutions of (\ref{DG}) can be put in the form (\ref{schwDG}). 
Given the solution (\ref{DG}), the temperature, specific heat, and entropy are given by \cite{Anninos:2017hhn, Witten:2020ert}
\begin{equation}\label{thermo}
T = \frac{U(\Phi(r_h))}{4\pi \tilde{\Phi} \ell}~, \quad\quad C = \frac{2\pi}{\kappa} \frac{U(\Phi(r_h))}{\partial_\Phi U(\Phi(r_h))}~, \quad\quad  S =  \frac{2\pi\Phi_0}{\kappa} + \frac{2\pi \tilde{\Phi}}{\kappa} r_h~.
\end{equation}
A non-negative entropy requires $\tilde{\Phi} \, r_h \ge -\Phi_0$, and a positive specific heat requires $\partial_\Phi U(\Phi(r_h))>0$.  $U(\Phi)$ must be positive for the metric to be positive definite. The partition function is given by
\begin{equation}
\log Z = S - \frac{E}{T}~, \quad\quad E = -\frac{1}{2\ell \kappa}\int_{r_h}^{r_b} d\tilde{r} \, U(\Phi(\tilde{r}))~,
\end{equation} 
where $r_b$ is the location of the Euclidean AdS$_2$ boundary. The energy diverges in the limit $r_b \to \infty$. This is a divergence can be absorbed in an infinite shift of the ground state energy. More meaningful is the energy difference between two solutions
\begin{equation}
E_{r_h} - E_{r_h'} = \frac{1}{2\ell \kappa}\int_{r_h'}^{r_h} d\tilde{r} \, U(\Phi(\tilde{r}))~.
\end{equation}
We note that $dS = T dE$, as expected. 

From the perspective of low energy effective field theory, the shape of the dilaton potential $U(\Phi)$ can be rather arbitrary. However, certain features are unacceptable at least from the perspective of thermodynamics. We assume that the form of $U(\Phi)$ complies with the positivity of entropy and gives rise to at least one solution with positive specific heat. If follows from (\ref{schwDG}) that a dilaton potential admitting asymptotically (near) AdS$_2$ geometries must be dominated by a linear behaviour in $\Phi$ in the asymptotic region.

\subsection{Adding matter fields} \label{adding_matter}

We can further imagine adding matter fields to our two-dimensional model. For this section it will be convenient to go to the Weyl gauge for the two-dimensional metric, namely
\begin{equation}
\frac{ds^2}{\ell^2} = e^{2\alpha(t,z)} \left( dt^2 + dz^2 \right)~, \quad\quad \Phi = \Phi(t,z)~,
\end{equation}
where $t$ can either lie on the whole real line or the circle depending on the properties of $\alpha(t,z)$. So long as backreaction of the matter fields is negligible, we can choose coordinates for which the background solution for the dilaton and two-dimensional metric are time-independent. In what follows, we consider matter that does not backreact and take $\Phi(t,z)=\Phi(z)$ as well as $\alpha(t,z)=\alpha(z)$. The relation to the Schwarzschild coordinates is given by
\begin{equation}\label{rtoz}
\Phi(z) =  \tilde{\Phi} \, r(z) =  \tilde{\Phi} \int^\infty_z d\tilde{z} \, e^{2\alpha(\tilde{z})}~, \quad\quad f(r(z)) = e^{2\alpha(z)}~.
\end{equation}
Let us consider a scalar field. At the quadratic level the most general two derivative covariant action is given by
\begin{equation}\label{2dmatter}
S_{\text{matter}} = \frac{1}{2} \int dt dz  \left( \mathcal{W}(\Phi)  \left( \partial_t \zeta  \partial_t \zeta +  \partial_z \zeta  \partial_z \zeta \right) + e^{2\alpha(t,z)} \mathcal{V}(\Phi)\zeta^2 \right)~.
\end{equation}
From the higher-dimensional perspective discussed in the previous sections, the functions $\mathcal{W}(\Phi)$ and $\mathcal{V}(\Phi)$ encode the behavior of $\gamma(t,z)$ and $\upsilon(t,z)$ in (\ref{4dconf}). The wave-equation governing $\zeta$ is now given by
\begin{equation} \label{scalar_eom}
 \left( - {\partial_z \mathcal{W}(\Phi(z)) \partial_z} + {\mathcal{W}(\Phi(z))}   \omega^2 +  \mathcal{V}(\Phi(z))  e^{2\alpha(z)}  \right) \zeta_\omega(z) = 0~.
\end{equation}
We will not impose particularly restrictive conditions on the form of $\mathcal{W}(\Phi(z))$ and $\mathcal{V}(\Phi(z))$.\footnote{One condition that may be natural to consider is that in the regime where the dilaton potential $U(\Phi)$ goes linearly in $\Phi$ up to small corrections, both $\mathcal{W}(\Phi(z))$ and $\mathcal{V}(\Phi(z))$ are constant up to comparably small corrections. From the higher-dimensional perspective, this corresponds to the dimensional reduction of a scalar field minimally coupled to a background which is Euclidean AdS$_2 \times S^2$ up to small deviations.} Thus we see that much of our discussion in the previous section translates directly to an analysis of the above equation. The map from $\upsilon(z)$ and $\gamma(z)$ in (\ref{probe4d}) to the two-dimensional problem is
\begin{equation}
e^{2\upsilon(z)} = \mathcal{W}(\Phi(z))~, \quad\quad m^2 e^{2\gamma(z)} =    \frac{\mathcal{V}(\Phi(z))}{ \mathcal{W}(\Phi(z))} \, e^{2\alpha(z)}~.
\end{equation}
We can ask what information is encoded in the boundary two-point function of $\zeta$. Provided we have knowledge of $\alpha(z)$, we can use (\ref{inverseLap}) and (\ref{upsilonR}) to go from the boundary correlation function to $\mathcal{V}(\Phi(z))$ and $\mathcal{W}(\Phi(z))$. This could allow us to test, for instance, whether the matter content corresponds to minimally coupled matter stemming from a four-dimensional theory. 

If a microscopic model is holographically dual to a dilaton-gravity theory of the type (\ref{DG}) we can construct $\alpha(z)$ from $U(\Phi)$ combined with (\ref{schwDG}) and (\ref{rtoz}). The dilaton potential $U(\Phi)$ can be constructed from the entropy as a function of the temperature, as follows from (\ref{thermo}). The presence of a weakly coupled nearly marginal scalar operator in the microscopic model  allows us to further construct the matter couplings $\mathcal{V}(\Phi)$ and $\mathcal{W}(\Phi)$ using the results of section \ref{sec3}. In the next section,  we study a microscopic model which has some of these features.
\newline\newline
\textbf{Example of a dimensional reduction.} As an example, the dimensional reduction of Einstein-Maxwell theory with four-dimensional metric
\begin{equation}\label{4dDG}
ds^2 = \frac{1}{\sqrt{\Phi}} g_{\mu\nu} dx^\mu dx^\nu + 4\Phi d\Omega_2^2~, \quad\quad \Phi \ge 0~,
\end{equation}
leads to a two-dimensional theory is of the form (\ref{DG}) with $U(\Phi) = (2\sqrt{\Phi})^{-1}$ plus a term involving a two-dimensional $U(1)$ gauge field. If our matter field stems from the dimensional reduction of a minimally coupled scalar field in four-dimensions, it follows \cite{Cavaglia:1998xj} from the parameterisation (\ref{4dDG}) that
\begin{equation}\label{minimallycoupled}
\mathcal{W}(\Phi) = 16\pi \Phi~, \quad\quad \mathcal{V}(\Phi) = 16\pi  m^2 \sqrt{\Phi}~,
\end{equation}
where $m$ is the mass of the particle. In the near horizon limit, where the geometry becomes AdS$_2\times S^2$, $\Phi $ is constant. Deviating slightly away from this limit, we have that $\Phi$ will be a constant up to small deviations. 

\section{Microscopic toy model, deformed.}\label{sec5}

In this section we discuss a quantum mechanical SYK-type model  \cite{Sachdev:1992fk, kitaev_vid,Kitaev:2017awl,Maldacena:2016hyu} built from a large number $N \in 2 \mathbb{Z}^+$ of fermionic degrees of freedom with random interactions. After introducing the model and discussing some of its features we will proceed to deform it by adding a relevant deformation. The undeformed model we consider has a low energy conformal phase with a near marginal operator in its spectrum. 

\subsection{SYK with two flavours}

The model we study is one of the models introduced and solved in \cite{Gross:2016kjj}. It is described by the following Hamiltonian
\begin{equation}\label{genSYK}
H_{q} = \frac{i^{q}}{(q!)^2} \sum_{\substack{i_1,\dots,i_q=1 \\ j_1,\dots,j_q=1}}^{N/2} J_{i_1,\ldots, i_q;j_1,\ldots, j_q} \psi_{i_1} \ldots \psi_{i_q}  \xi_{j_1} \ldots \xi_{j_q}~, 
\end{equation}
where the $N/2$ real fermions $\psi_i$ and the $N/2$ real fermions $\xi_j$ obey the anti-commutation relations $\{ \psi_i ,\psi_j \} = \delta_{ij}$, $\{ \xi_i ,\xi_j \} = \delta_{ij}$, and $\{ \xi_i,\psi_j\} = 0$. The number of fermions $q \in 2\mathbb{Z}^+$ in the interaction term is even, but we will further take it to be multiple of four for future convenience. The couplings are taken to be random, coming from a Gaussian ensemble with variance
\begin{equation} \label{gaussian}
\langle J_{i_1,\ldots, i_q;j_1,\ldots, j_q} J_{i_1,\ldots, i_q;j_1,\ldots, j_q} \rangle  = \frac{J_{q}^2}{N^{2q-1}} \frac{(q-1)!^2}{2^{2q}} \,.
\end{equation}
They are totally anti-symmetric in each set of indices $\{i_1,\ldots,i_q\}$ and $\{j_1,\ldots,j_q\}$. To leading order in the large $N$ expansion, the model exhibits a low temperature phase with an approximate time-reparameterisation invariance. At zero temperature and for $J_q \, t \gg 1$, the large $N$ Euclidean two-point function is given by 
\begin{equation}
G_\psi(t) \equiv \frac{2}{N} \sum_{i=1}^{N/2} \langle \psi_i (t) \psi_i(0) \rangle = \frac{2}{N} \sum_{i=1}^{N/2}  \langle \xi_i (t) \xi_i(0) \rangle = b_{\Delta_\psi} \frac{\text{sgn} \, t}{|J_q  t |^{2\Delta_\psi}} ~,
\end{equation}
with $\Delta_\psi =1/2q$. The normalisation constant $b_{\Delta_\psi}$ is computed in \cite{Gross:2016kjj} $b_{\Delta_\psi}$ for arbitrary SYK models with flavours. For the model we are considering, one finds $b_{\Delta_\psi} = 2^{-3 \Delta_\psi}$  in the large-$q$ limit. We further note that $G_\psi(-t) = -G_\psi(t)$.

The spectrum of conformal operators in the low temperature phase includes a tower of increasing weights. We will focus on the following Hermitean operator
\begin{equation}\label{Ozeta}
\mathcal{O}_\zeta = \frac{2i}{N} \sum_{i=1}^{N/2}   \psi_i \xi_i~.
\end{equation}
We would like to calculate the two-point function of $\mathcal{O}_\zeta$ and establish its scaling dimension in the low temperature conformal phase. A straightforward examination of the contributing Feynman diagrams indicates that to order $1/N$ only the disconnected diagram contributes. Thus, 
\begin{equation} \label{G_scalar}
G_{\zeta}(t) \equiv \langle \mathcal{O}_\zeta(t) \mathcal{O}_\zeta(0) \rangle = \frac{1}{N} G_\psi(t)^2 + O(N^{-2}) = \frac{1}{N} \frac{b_{\Delta_\zeta}}{| J_q t|^{2\Delta_\zeta}}~,
\end{equation}
and we note that $G_{\zeta}(t) = G_{\zeta}(-t)$. The conformal dimension of $\mathcal{O}_\zeta$ is $\Delta_\zeta = 2\Delta_\psi$. When $\Delta_\psi$ is small, so is the scaling dimension of $\mathcal{O}_\zeta$. 

In order to connect to the discussion in section \ref{sec3}, we need a scalar operator with scaling dimension parameterically close to $1$. For this, we can use the results of \cite{Klebanov:1999tb, Mueck:2002gm, Gubser:2002vv}. In those works, it is shown that to leading order in the large $N$ limit, a $d$-dimensional conformal field theory deformed by a double trace deformation $\mathcal{O}_\Delta^2$ with coupling $f$ results in a theory which is also conformal, but one in which the operator $\mathcal{O}_\Delta$ has conformal weight $\Delta' = d-\Delta$. Applied to (\ref{genSYK}), by adding a double-trace deformation $\mathcal{O}_\zeta^2$ to the theory we can obtain an operator of weight $\Delta_\zeta' = 1-\Delta_\zeta$ which becomes parameterically close to one as we take $q$ to be large. 

The thermodynamic properties of (\ref{genSYK}) can also be computed. If we first take the large $N$ limit and then take the small temperature limit the model has an extensive ground state degeneracy and a specific heat linear in the temperature. Recalling (\ref{thermo}) and assuming one can associate a two-dimensional dilaton-gravity model to the low temperature conformal phase, we see that a linear in temperature specific heat corresponds to a dilaton potential of the form $U(\Phi)  \propto  \Phi$. This yields a Euclidean AdS$_2$ metric since the Ricci scalar is given by $R = - \partial_\Phi U(\Phi)$. That the two-point functions of the conformal operators take the standard scale-covariant form indicate that $\mathcal{W}(\Phi)$ and $\mathcal{V}(\Phi)$ in (\ref{2dmatter}) are constant in the conformal phase. 

\subsection{An exact RG flow}

Having introduced the model (\ref{genSYK}) with a low temperature conformal phase, we would now like to turn on a relevant deformation that induces an RG flow. The relevant deformation we consider takes the form of an additional Hamiltonian $H_{\tilde{q}}$ of the type (\ref{genSYK}) with $q$ replaced with $\tilde{q}$:
\begin{equation}\label{Hdeformed}
H_{\text{def}} = H_{q} + s \, H_{\tilde{q}}~.
\end{equation}
Here $s$ is a tuneable parameter which we take to be positive without loss of generality. 
Remarkably, for large values of $q$ and $\tilde{q}$ the above theory is still solvable \cite{Jiang:2019pam}. In what follows we take $q/\tilde{q}$ to be a fixed parameter as $q$ becomes large, and further that $q$ does not scale with $N$ in the large $N$ limit. We will also assume that $s$ does not scale  with $q$ or $N$ and that $\tilde{q}$ is even. For simplicity we will restrict ourselves to the case $\tilde{q}=q/2$.\footnote{The reason for choosing $\tilde{q}<q$, and hence $2\tilde{q}\Delta_\psi<1$, is that we would like our deformation to correspond to a relevant deformation of the IR fixed point. Since we are deforming the UV Hamiltonian, this is not guaranteed. However, as we shall soon see, it will indeed be the case for certain regimes of parameter space.}

We begin by discussing the fermion two-point function along the flow. This is obtained by solving the following Schwinger-Dyson equations
\begin{eqnarray}
\Sigma_\psi & = & s^2 \frac{4J_{q/2}^2}{q} G_\psi^{q-1} + \frac{2J_{q}^2}{q} G_\psi^{2q-1} \,, \label{SD1}\\
G_\psi & = & \left(  \partial_{t} - \Sigma_\psi \right)^{-1}  \,, \label{SD2}
\end{eqnarray}
where $\Sigma_\psi$ is the usual self-energy that appears in the SYK model. Some details on the derivation are provided in appendix \ref{F_app}.
In the large-$q$ limit, we parameterise $G_\psi(t)$ for $t\neq0$ as 
\begin{equation}\label{Gfermion}
G_\psi(t)  = \frac{ {\text{sgn}} \, t }{2} \left(1 + \frac{g(t)}{q} + \cdots \right) \,.
\end{equation}
We also have that $G_\psi(t=0)=0$. Then, to leading order in the large-$q$ expansion, the Schwinger-Dyson equations reduce to \cite{Jiang:2019pam}
\begin{equation}
\partial_t^2 g(t) = 2 s^2 {\cal{J}}_{q/2}^2 e^{g(t)} + {\cal{J}}_q^2 e^{2g(t)}\,, \label{g_eq}
\end{equation}
where we conveniently defined ${\cal{J}}_x^2 \equiv 2^{3-2x} J^2_{x}$. Let's further assume that ${\cal{J}}_q = {\cal{J}}_{q/2} \equiv {\cal{J}}$. This equation can be solved exactly for any value of the couplings, obtaining
\begin{equation}\label{expg}
e^{g(t)} = \frac{1}{1+ {\cal{J}} |t| \sqrt{ 1 +4  s^2}+ s^2 {\cal{J}}^2 |t|^2} \,, \quad\quad t\in\mathbb{R}~.
\end{equation}
The boundary conditions are chosen so that $e^{g(0)} =1$ and $e^{g(\infty)} = 0$. Consequently, the two-point function of the scalar operator $\mathcal{O}_\zeta$ becomes
\begin{eqnarray} \label{two_point_q}
G_{\zeta}(t) = \frac{1}{4}+\frac{1}{2 q} \log \left( \frac{1}{1+ {\cal{J}} |t| \sqrt{ 1 +4  s^2}+ s^2 {\cal{J}}^2 |t|^2} \right) + O\left( q^{-2} \right) \,.
\end{eqnarray}
For $s=0$, we recover the two-point function of the undeformed model (\ref{genSYK}) in the large-$q$ limit. For $s \gg 1$, the late-time two-point function becomes that of an undeformed model with Hamiltonian $s \, H_{\tilde{q}}$ at large $\tilde{q}$. For $0<s \ll 1$, the late-time two-point function interpolates between one conformal behaviour ($1\ll {\cal{J}}|t| \ll 1/s^2$) to another ($1/s^2 \ll {\cal{J}}|t|$). Thus, provided $s$ is sufficiently small we can view the deformation as a relevant deformation causing the original (near) conformal IR fixed point to flow to another. We will explore the finite temperature manifestation of this RG flow in section \ref{thermalRG}.

It is also instructive to study the two-point function in frequency space. The leading term in (\ref{two_point_q}) is a constant that can be interpreted as an expectation value, so we will be interested in computing the Fourier transform of the difference $\delta G_\zeta (t) = G_\zeta (t)- 1/4$. Since we only expect to make a connection with a dual gravitational description in the strongly coupled IR phase, we will work in the limit of $\mathcal{J} |t| \gg 1$. To leading order
\begin{equation} \label{two_p_time}
\delta G_{\zeta}(\tilde{t}) = -\frac{1}{2q}  \log \left( \sqrt{1+4s^2} \, |\tilde{t}| + s^2 |\tilde{t}|^2 \right) + \ldots \,,
\end{equation}
where we defined the dimensionless time $\tilde{t} = \mathcal{J} t$. The Fourier transform of $\delta G_{\zeta}(\tilde{t})$ with respect to $\tilde{t}$ is given by
\begin{equation}
\delta G_{\zeta}(\tilde{\omega}) =  \frac{1}{{2 q \sqrt{2 \pi } | \tilde{\omega} | }}\left(2 \text{Ci}\left(| \tilde{\omega} |/\aleph_s\right) \sin \left(| \tilde{\omega} |/\aleph_s\right)+\left(\pi -2 \text{Si}\left(| \tilde{\omega} |/\aleph_s \right)\right) \cos \left(| \tilde{\omega} |/\aleph_s\right)+\pi \right) + \cdots\,,
\end{equation}
where $\tilde{\omega} = \omega/{\cal{J}}$, ${\text{Ci}}(x)$ and ${\text{Si}}(x)$ are the cosine and sine integral functions respectively, and
\begin{equation} \label{aleph_def}
\aleph_s \equiv \frac{s^2}{\sqrt{1+4s^2}}~.
\end{equation}
Note that $\aleph_s$ is greater or equal than zero. For $s^2 \ll 1$, $\aleph_s \approx s^2$, while for $s^2 \gg1, \aleph_s \approx s/2$. 

As already mentioned, to make contact with the discussion in section \ref{sec3}, we need a two-point function of an operator with dimension close to one in the (near) conformal phase. To achieve this we come back to the discussion in \cite{Mueck:2002gm, Gubser:2002vv}. Turning on the composite operator $\mathcal{O}_\zeta^2$ with coupling $f$ we obtain a new large $N$ two-point function
\begin{equation} \label{GK_trick}
Q_{\zeta}(\tilde{\omega}) = \frac{\delta G_{\zeta}(\tilde{\omega})}{1+ f \delta G_{\zeta}(\tilde{\omega})}~.
\end{equation}
In the limit $f \delta G_{\zeta}(\tilde{\omega}) \gg 1$ we find
\begin{equation} \label{Q_zeta}
Q_{\zeta}(\tilde{\omega}) = - \frac{1}{f^2 \delta G_\zeta(\tilde{\omega})} + \text{local}~.
\end{equation}
First consider $\aleph_s = 0$. To leading order in the large-$q$ expansion, we have
\begin{equation}
\left. Q_{\zeta}(\tilde{\omega}) \right|_{\aleph_s = 0} = -\frac{2q}{f^2} \sqrt{\frac{2}{\pi}} |\tilde{\omega} | \,,
\end{equation}
which is just the conformal two-point function for the $\Delta=1$ scalar $\mathcal{O}_\zeta$ in the two-flavoured SYK model with a $2q$-interaction, as expected. 
For $1 \gg \aleph_s > 0$, we can obtain analytic expressions for $\tilde{\omega} \ll \aleph_s$ and $\tilde{\omega} \gg \aleph_s$:
\begin{eqnarray}
\text{small}\, \tilde{\omega}: & Q_{\zeta}(\tilde{\omega})  =   -\frac{q \, \aleph_s}{f^2} \sqrt{\frac{2}{\pi }} \left( \frac{\left| \tilde{\omega }\right|}{\aleph_s} -\frac{1} {\pi} \frac{\left| \tilde{\omega }\right| ^2 \log \left| \tilde{\omega } \right|}{\aleph_s^2}  +O \left( \left( {\left| \tilde{\omega} \right|}/{\aleph_s} \right) ^3 \right) + \text{local} \right)  +  \cdots \,, \label{smallw}  \\
\text{large}\, \tilde{\omega}:  & Q_{\zeta}(\tilde{\omega})  =  -\frac{q \, \aleph_s}{f^2} \sqrt{\frac{2}{\pi }}\left(   \frac{2\left| \tilde{\omega }\right| }{\aleph_s }-\frac{4}{\pi }+\frac{8}{\pi ^2} \frac{\aleph_s}{\left| \tilde{\omega }\right| } + \frac{8 \left(\pi ^2-2\right)}{\pi ^3} \frac{\aleph_s^2}{\left| \tilde{\omega }\right|^2 }+O\left(\left({\aleph_s}/{\left| \tilde{\omega} \right|} \right)^{3}\right) \right)  + \ldots  \label{largew}
\end{eqnarray}
Note that both for small and large $\tilde{\omega}$, the propagator is proportional to $|\tilde{\omega}|$, that is the conformal propagator for $\Delta =1$. Moreover, the relative coefficient does not depend on $\aleph_s$ and is exactly 2, a direct consequence of the theory flowing from a $q$-model for small $\tilde{\omega}$ to $2q$ for large $\tilde{\omega}$. It is important to stress that while the large $\tilde{\omega}$ expansion is analytic in $\tilde{\omega}$, the small $\tilde{\omega}$ expansion contains logarithms.

\subsection{Thermodynamics}\label{thermalRG}
In this section we study the thermodynamics of the model, that will give us access to the dilaton potential in the bulk theory. It is possible to also compute the exact form of the two-point function at finite temperature, that is given by solving (\ref{g_eq}) with boundary conditions $g(0) = g(\beta) = 0$. The result is given by
\begin{equation} \label{g_finite_T}
e^{g(\tau)} = \frac{2 \nu ^2}{\sqrt{(\beta {\cal{J}})^2 \nu ^2+ s^4 (\beta  {\cal{J}})^4} \cos (\nu  (\frac{2 \tau}{\beta}-1))+ s^2 (\beta {\cal{J}})^2}
\end{equation}
with
\begin{equation}\label{nuequation}
\cos \nu = \frac{2 \nu ^2- s^2 (\beta {\cal{J}})^2}{\sqrt{(\beta {\cal{J}})^2 \nu ^2+ s^4 (\beta {\cal{J}})^4}} \,.
\end{equation}
Here, $\tau \sim \tau+\beta$ is the periodic Euclidean time variable. The equation (\ref{nuequation}) has multiple solutions. We pick the one continuously connected to $\nu = 0$ at $\beta \mathcal{J} = 0$. For this solution, $\nu = \pi$ at $\beta \mathcal{J} \to \infty$. The thermal two-point function is then
\begin{equation}
G_\psi(\tau)  = \frac{ {\text{sgn}} \, \tau }{2} \left(1 + \frac{g(\tau)}{q} + \cdots \right) \,.
\end{equation}
Having obtained an expression for the thermal two-point function along the flow, we can compute the thermodynamic variables of the deformed model. This was elegantly done for a related model with no flavours in \cite{Jiang:2019pam}, and the derivation is essentially the same for the model under consideration. We simply state the answer, providing some details and numerical checks in appendix \ref{F_app}. To leading order in the large-$q$ limit, the thermodynamic quantities can be succinctly expressed implicitly in terms of a thermodynamic variable $x$ that ranges from 0 to 1. The inverse temperature is given by
\begin{equation} \label{temperature}
\beta {\cal{J}} = \frac{2 \Theta(x)}{ \sqrt{x (4 s^2+x)}} \,\,\,\, \text{with} \,\,\,\,\, \cos \Theta(x) = \frac{s^2 (4 x-2)+x^2}{2 s^2+x} \,.
\end{equation}
Note that $x=0$ implies $\beta {\cal{J}}= \infty$ while $\beta {\cal{J}} = 0$ at $x=1$. The thermal partition function is given by
\begin{multline}\label{Zexact}
\frac{1}{N} \log Z_{\text{thermal}}(x) = \frac{S_0}{N} + \frac{\Theta(x)^2}{4 q^2} + \frac{\beta {\cal{J}}}{4q^2} \left(\sqrt{(1-x) (4 s^2+x+1)}\, + \right. \\ \left. 4 s^2 \sinh ^{-1}\sqrt{\frac{1-x}{4 s^2+2 x}}
-2 \sqrt{x (4 s^2+x)} \tan ^{-1}\sqrt{\frac{(1-x) (4 s^2+x)}{x (4 s^2+x+1)}} \right)  \,,
\end{multline}
where $S_0 = \frac{N}{2}\log 2$ is the entropy of $N$ free fermions. Note that this is exactly the same as the thermal partition function of a model with $q$ and $2q$ terms, without any flavours \cite{Jiang:2019pam}. 

The entropy can also be computed, yielding the much simpler expression
\begin{eqnarray}
S (x) & = & S_0 - \frac{N \Theta(x)^2}{4q^2} \,. \label{entropy}
\end{eqnarray}
\begin{figure}[h!]
        \centering
                \includegraphics[scale=0.7]{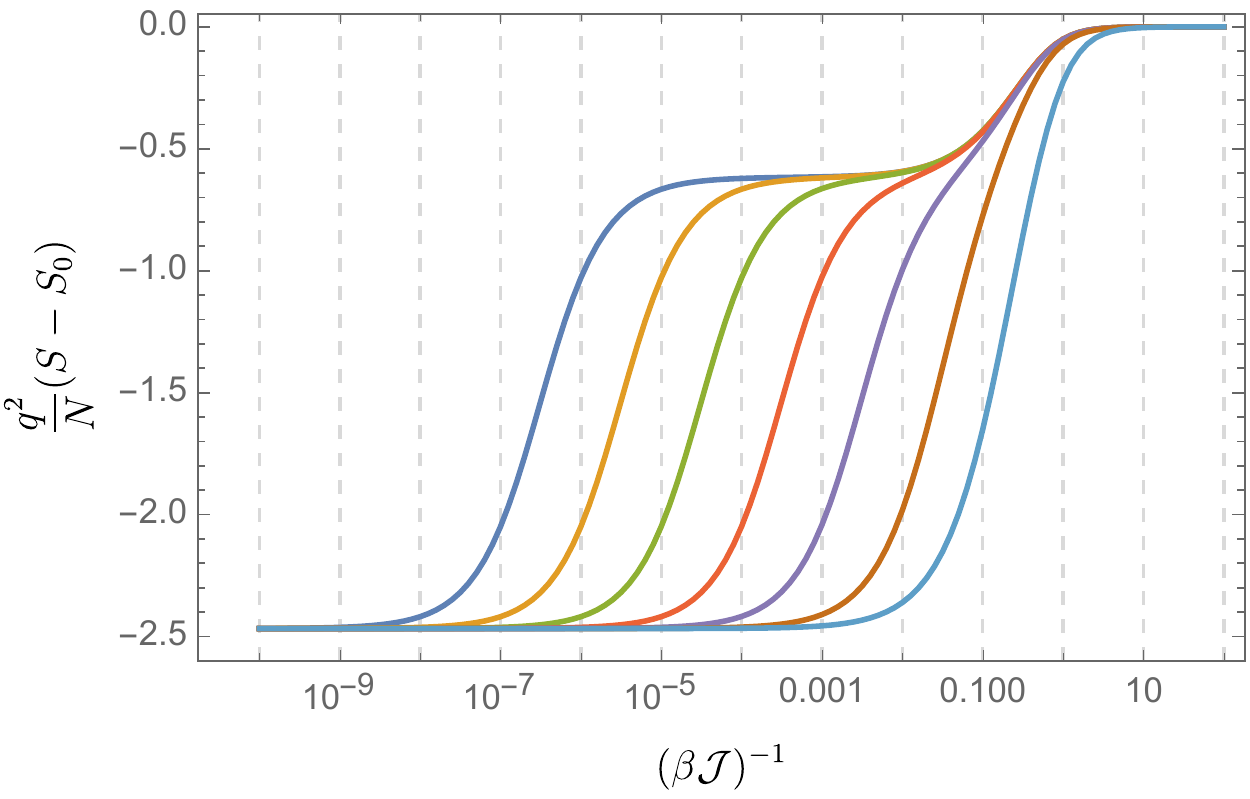}
                                 \caption{{\footnotesize Entropy as a function of the temperature (in logarithmic scale) to leading order in the large $N$ and $q$ expansion. Different curves correspond to different values of $s^2=10^{-6},10^{-5}, 10^{-4},10^{-3}, 10^{-2}, 10^{-1}, 10^{0}$, from left to right.  }}
\label{fig_entropy}
\end{figure}
\begin{figure}[h!]
        \centering
        \subfigure[Deep IR]{
                \includegraphics[scale=0.58]{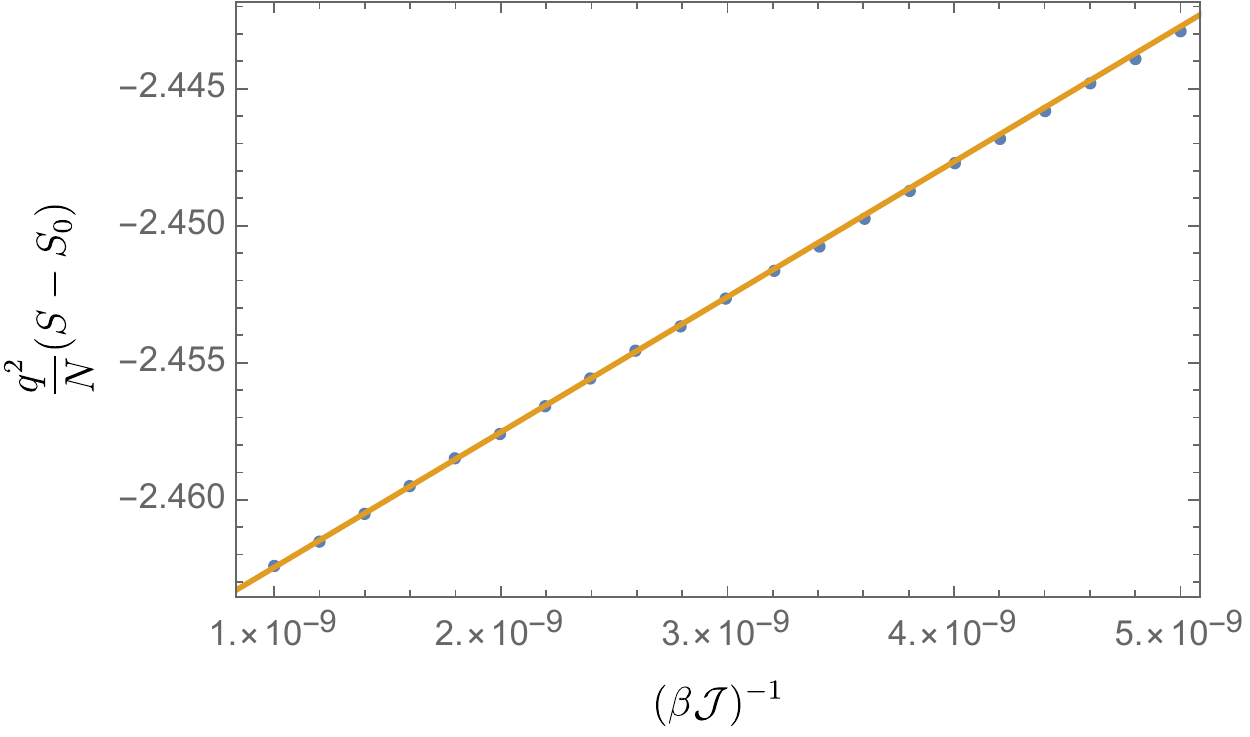} \label{fig_deep}} \quad\quad
         \subfigure[Intermediate IR]{
                \includegraphics[scale=0.58]{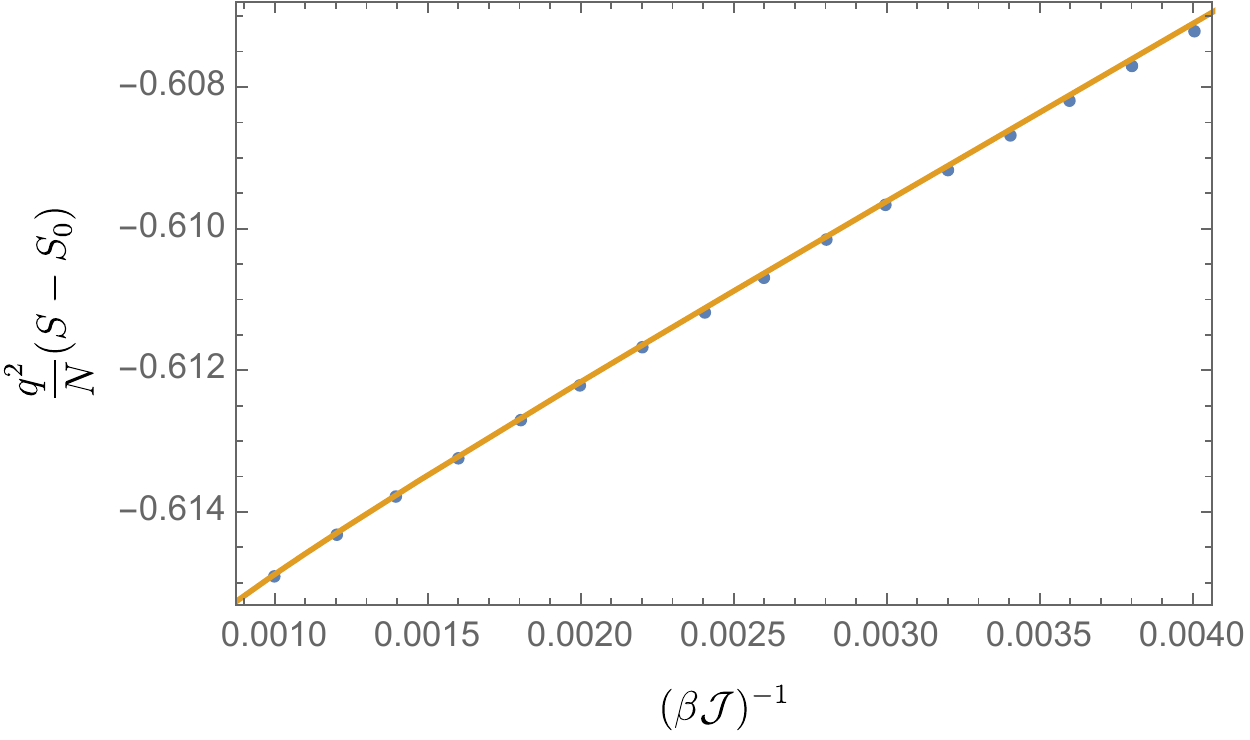} \label{fig_int}}
                 \caption{{\footnotesize Entropy as a function of the temperature in the two linear regimes. The blue dots correspond to the exact evaluation of (\ref{entropy}), while the solid yellow lines correspond to the approximations in (\ref{series_deepIR}) and (\ref{series_intIR}). In both cases we fix $s^2=10^{-6}$. }}
\label{fig_linearentropy}
\end{figure}
Given equations (\ref{temperature}) and (\ref{entropy}), we can implicitly plot the entropy as a function of temperature.  This is shown for different values of $s^2$ in figure \ref{fig_entropy}. In the large temperature regime, where $\beta {\cal{J}} \ll 1$ while keeping $s^2$ fixed the entropy goes to $S_0$. The interesting behaviour comes as $\beta {\cal{J}}$ grows. We see that for any fixed $s^2$, at sufficiently low temperatures all curves reach a ``lower plateau''. In fact, this is a regime of entropy being proportional to the temperature, as can be seen when we zoom into that region, as displayed in figure \ref{fig_deep}. For sufficiently small $s^2$, a second intermediate plateau appears, that also corresponds to entropy linear in temperature --- see figure \ref{fig_int}. Both regimes can be found analytically by studying (\ref{entropy}). Recall that in the ordinary SYK model with a large-$q$ interaction the entropy for $\beta \J \gg 1$ goes as \cite{Maldacena:2016hyu}
\begin{eqnarray}
{\text{SYK with $q$-interaction:}} \,\,\,\,\,  S - S_0 = \frac{N}{q^2} \left( -\frac{\pi^2}{4}+\frac{\pi ^2}{\beta {\cal{J}}} + \cdots \right) \,. \label{series_SYK_deepIR}
\end{eqnarray}
Consider now the deformed model in the very small temperature regime. Expanding (\ref{entropy}) in large $\beta \mathcal{J}$, with $\beta \mathcal{J} \gg 1/\aleph_s$, and using (\ref{temperature}) we obtain,
\begin{eqnarray}
{\text{Deep IR:}} \,\,\,\,\,  S-S_0  =  \frac{N}{q^2}  \left(-\frac{\pi ^2}{4}+ \frac{\pi^2}{2 \aleph_s} \frac{1}{\beta \mathcal{J}} + \ldots \right)~.
\label{series_deepIR}
\end{eqnarray}
The coefficient linear in the temperature depends on $\aleph_s$, as defined in (\ref{aleph_def}). For $\aleph_s \gg 1$ the above formula reduces exactly to the entropy of the underformed SYK model with a large-$q$ interaction and coupling $s \mathcal{J}$. Now consider the case where $\aleph_s \ll 1$. The deep IR region occurs for temperatures satisfying $\beta \J \gg 1/\aleph_s $. In this case, the entropy becomes again (\ref{series_deepIR}). On the other hand, when $1 \ll \beta \J \ll 1/\aleph_s$ a new intermediate IR region emerges. This is exhibited in figure \ref{fig_entropy}. In the regime $1 \ll \beta \J \ll 1/\aleph_s$ the entropy goes as
\begin{eqnarray}\label{Sinter}
{\text{Intermediate IR:}} \,\,\,\,\, S-S_0  = \frac{N}{q^2}  \left(  -\frac{\pi ^2}{16} + \frac{\pi^2}{4\beta {\cal{J}}}  -\frac{\aleph_s \beta {\cal{J}}}{2} + \aleph_s + \cdots \right) \,. \label{series_intIR}
\end{eqnarray}
Note that the term independent of the temperature is the one of the undeformed SYK with a $2q$ interaction plus a small correction. The linear in temperature coefficient also agrees with the coefficient in standard $2q$-SYK. Interestingly, in contrast to (\ref{series_deepIR}), this term does not depend on $\aleph_s$. When $\aleph_s$ goes to zero, only this intermediate regime survives and we recover exactly the thermodynamics of the $2q$-SYK model. Note that in order for the linear-in-temperature term to dominate in (\ref{series_intIR}), we actually further need $\beta \J \ll 1/\sqrt{\aleph_s}$. These expansions can be checked numerically against the actual curve and we find good agreement, as shown in figure \ref{fig_linearentropy}. 

It is straightforward to compute the specific heat $C = - \beta \partial_\beta S$.
When $S$ is linear in the temperature, so $C$. In particular in our deformed model, we find that
\begin{eqnarray}
{\text{Deep IR $(\beta \J \gg \text{max} (1,1/\aleph_s))$:}} & & C \approx \frac{\pi^2}{2\aleph_s} \frac{N}{q^2 \beta  {\cal{J}}} \,,  \\ \label{Cmicro1}
{\text{Intermediate IR $(1 \ll \beta \J \ll 1/\sqrt{\aleph_s})$:}} & & C \approx \pi^2 \frac{N}{(2q)^2 \beta  {\cal{J}} }  \,. \label{Cmicro2}
\end{eqnarray}
Given the above expressions, it may be tempting to analytically continue $\aleph_s$ away from $\mathbb{R}^+$ to negative values. This can be achieved by continuing $s$ to a regime where $-1/4<s^2<0$. In appendix \ref{F_app} we provide numerical evidence that $s$ can be continued to a complex number of non-vanishing real part. We will return to this question in future work.

\section{Dilaton-gravity plus matter, holographically.}\label{reconstructionsec}

In this section we assess some properties of a putative bulk theory dual to the low energy sector of the microscopic model (\ref{Hdeformed}). To do so, we apply the techniques and ideas developed in sections \ref{sec3} and \ref{sec_grav}.

\subsection{Dilaton potential, reconstructed}\label{dilpotrec}

We first extract the dilaton potential through thermodynamic considerations.
\newline\newline
\textbf{Case I: $s=0$.} In this case we have a specific heat which is linear in the temperature at low temperature. According to (\ref{thermo}) the dilaton potential is thus linear in $\Phi$, leading to a dilaton-gravity theory of the form
\begin{equation}
S_E = -\frac{1}{2\kappa} \int d^2x \sqrt{g} \Phi \left( R - \frac{2}{\ell^2} \right) - \frac{1}{\kappa}\int_{\partial\mathcal{M}} du \sqrt{h} \Phi_b K~,
\end{equation}
where we have reinstated $\ell$ as the AdS$_2$ length and the theory lives on a disk. Imposing Dirichlet boundary conditions, we fix the induced metric $h$ and dilaton profile $\Phi_b$ at the $S^1$ boundary of the disk which is parameterised by the coordinate $u \sim u + 2\pi \tilde{\beta}$.   
The classical saddle obeying our boundary condition is
\begin{equation}
\frac{ds^2}{\ell^2} = \frac{1}{\tilde{\beta}^2} \, \frac{d u^2+ dz^2}{\sinh^2 z/\tilde{\beta}}~, \quad\quad \Phi =  \frac{\tilde{\Phi}}{\tilde{\beta}} \coth \frac{z}{\tilde{\beta}}~,
\end{equation}
where the boundary of the disk lives at $z = z_c \ll \tilde{\beta}$, so $h =\ell^2/z_c^2$ and $\Phi_b =  \tilde{\Phi}/z_c$.  
 From the on-shell action for the above solution we find the specific heat \cite{Jensen:2016pah, Maldacena:2016upp}
\begin{equation}
C_{\text{bulk}} = 2 \pi \frac{\tilde{\Phi}}{\tilde{\beta}\kappa}~.
\end{equation}
Comparing to the microscopic expression (\ref{Cmicro2}) $C_{\text{micro}} = \pi^2 N T_{\text{micro}} /(2 q)^2\mathcal{J}$, we can identify
\begin{eqnarray} \label{ids}
\frac{\pi^2 N}{4q^2} = \frac{1}{\kappa} \,\,\,\, , \,\,\,\,  2\pi T_{\text{micro}} = \frac{1}{\ell\tilde{\beta}}~, \,\,\,\,\, \text{and} \,\,\,\,  \mathcal{J} = \frac{1}{\ell \tilde{\Phi}} \,,
 \end{eqnarray}
 such that $\tau_{\text{micro}} = \ell u$ is identified with the Euclidean time coordinate of the microscopic theory. At vanishing temperature, the solution becomes 
 \begin{equation}\label{zeroTDG}
 \frac{ds^2}{\ell^2} = \frac{d u^2+ dz^2}{z^2}~, \quad\quad \Phi =  \frac{\tilde{\Phi}}{z}~,
 \end{equation}
with $u \in \mathbb{R}$ and $t_{\text{micro}} = \ell u$. Note that the dilaton vanishes at $z = \infty$. The microscopic entropy at vanishing temperature can be macroscopically accommodated by a topological term
\begin{equation} \label{topological}
S_{\text{top}} = -\frac{\Phi_0}{2\kappa} \int d^2x \sqrt{g} R - \frac{\Phi_0}{\kappa}\int_{\partial\mathcal{M}} du \sqrt{h} K~,
\end{equation}
with $\Phi_0$ tuned to match the zero-temperature entropy. Comparing to (\ref{series_intIR}), this gives $\Phi_0 = q^2 \log 2/\pi^3 - 1/8\pi$. Thus, it follows that $U(\Phi=0) = 0$.
\newline\newline
\textbf{Case II: $s\neq0$.} Once again, we would like to employ (\ref{thermo}) to construct a dilaton-gravity model that reproduces the thermodynamic behaviour. Again, our hypothesis is that the thermodynamic behaviour in the strong coupling regime is captured entirely by a dilaton-gravity sector, but now with a non-linear $U(\Phi)$.\footnote{Recent considerations of deformed dilaton potentials in pure dilaton-gravity theory include \cite{Witten:2020wvy,Maxfield:2020ale}.} We will maintain the identifications (\ref{ids}) for this case as well. The zero temperature entropy when $s^2$ is non-vanishing is given by,
\begin{equation}
S_{\text{micro}}(T=0) = \frac{N}{2} \log 2 - \frac{N \pi^2}{4q^2} \,,
\end{equation}
that will be macroscopically matched to a topological term (\ref{topological}) with $\Phi_0 = q^2 \log 2/\pi^3 -1/2\pi$. This constant differs from the vanishing $s^2$ case since now the entropy in the deep IR is given by the entropy of a $q$-SYK model instead of a $2q$-model occurring for $s^2=0$.

\begin{figure}[h!]
        \centering
        \subfigure[Full dilaton potential]{
                \includegraphics[height=5cm]{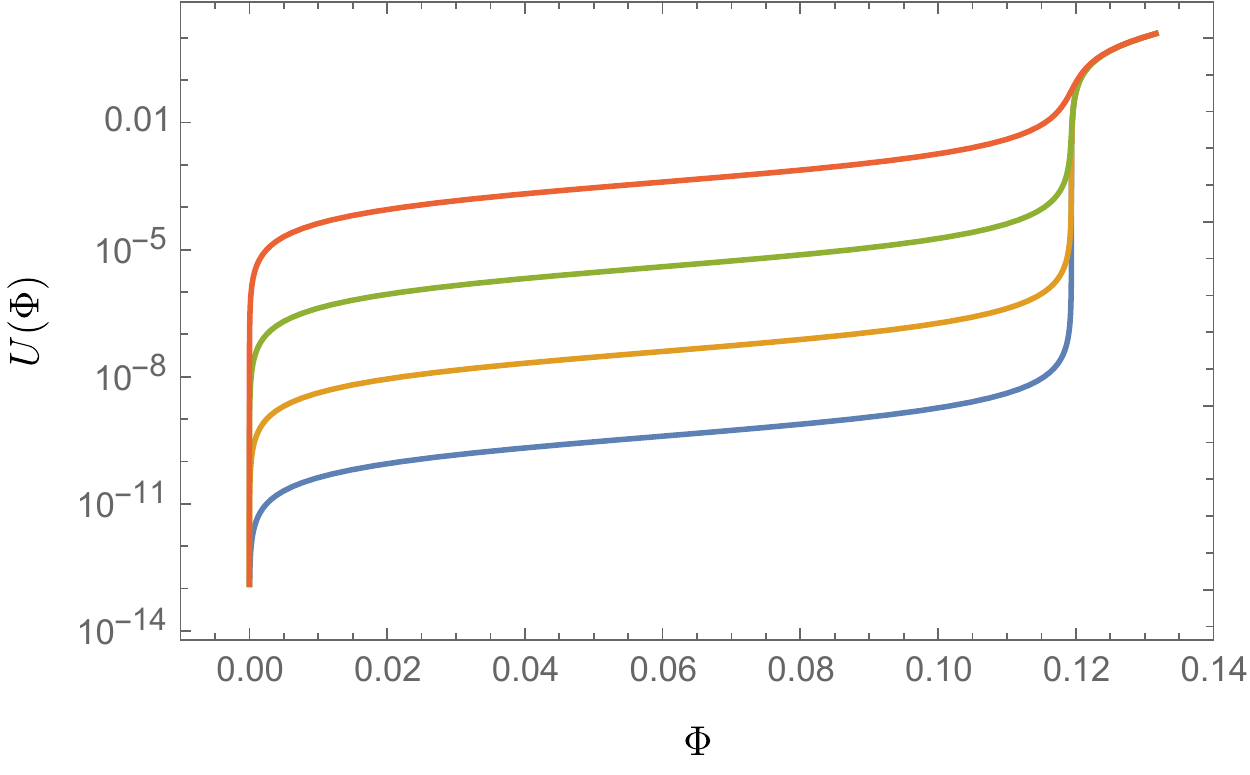} \label{fig_dil}} \quad\quad
         \subfigure[Close-to-linear dilaton potential]{
                \includegraphics[height=5.2cm]{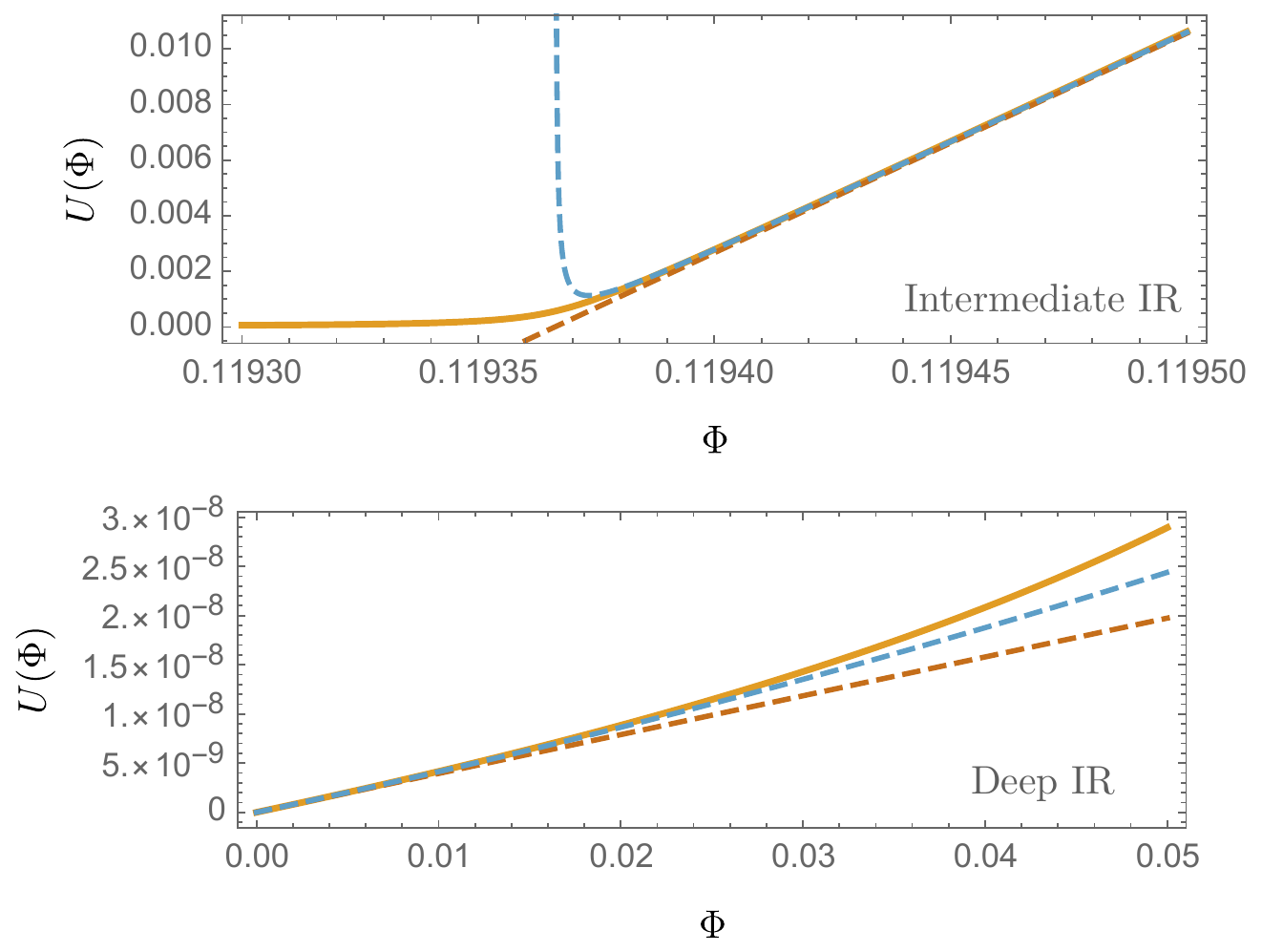} \label{fig_dil2}}
                 \caption{{\footnotesize (a) Dilaton potential for $s=10^{-5,-4,-3,-2}$, from bottom to top. The almost vertical regimes in this logarithmic plot correspond to regimes of linear dilaton potential, that we zoom in (b). In a physical setting, the potential will be cut-off in the UV at a value of $\Phi$ where the potential is still linear, so the curvy features in the upper-right corner of the plot will not be of interest. \newline
                 (b) We zoom close to the linear regimes in the deep and intermediate IR regions for the yellow curve with $s=10^{-4}$. The dashed lines correspond to the approximations given in (\ref{potential_deep}) and (\ref{potential_int}). The brown line corresponds to keeping only the linear term, while in light blue we also included the first correction. }}
\label{fig_linear}
\end{figure}

Once $\Phi_0$ is set, it is straightforward to obtain the dilaton potential by writing the temperature as a function of the entropy using (\ref{thermo}) and (\ref{ids}). We show a plot of the dilaton potential for different values of $s^2$ in figure \ref{fig_dil}.
The shape of $U(\Phi)$ for small values of $s$ indicates that the bulk metric interpolates between two AdS$_2$ regions, one near the boundary and one in the deep interior. There is also a non-vanishing dilaton profile. A horizon is still present in the deep interior, macroscopically encoding the microscopic entropy (\ref{series_deepIR}). Between the two AdS$_2$ regions lies an interpolating region which breaks the (near) $SL(2,\mathbb{R})$ isometry. None of the simple microscopic operators acquire an expectation value throughout the flow. From the bulk perspective, this translates to no matter fields being excited in the interpolating region -- the flow is captured entirely by the dilaton-gravity sector. 

We can understand some features of this potential analytically. Using the expressions for the entropy in the deep IR (\ref{series_deepIR}) and in the intermediate region (\ref{series_intIR}), we obtain
\begin{eqnarray}
\text{Deep IR}: &  &\,\,\,\,  U(\Phi) = 4 \pi ^2 \aleph_s \left( \Phi +  \frac{3 \pi }{2} \Phi^2 + \cdots \right)\,,  \label{potential_deep} \\
\text{Intermediate IR}: & &  \,\,\,\,  U(\Phi) = 8 \pi ^2 \left(\Phi -\frac{3}{8 \pi }\right) + \frac{4 \aleph_s}{\pi ^2} {\left(\Phi -\frac{3}{8 \pi }\right)^{-1}} + \cdots \label{potential_int} \,.
\end{eqnarray}
The expansion (\ref{potential_int}) is valid provided that $\sqrt{\aleph_s} \ll (\Phi-3/8\pi) \ll 1$ and $\aleph_s \ll 1$.

Note that when $\aleph_s=0$, which corresponds to $s^2=0$, the deep IR entropy vanishes and we are only left with the intermediate IR where the dilaton has a shift that matches the result in the previous subsection. Figure \ref{fig_dil2} shows agreement between these expressions and the full potential. In each region, we included the linear term and its first correction. From this, we can extract the metric in the Schwarzschild gauge via the relation (\ref{schwDG}). Finally, we can go to the conformal gauge via (\ref{rtoz}) leading us to an expression for $\Phi(z)$ and $\alpha(z)$. We can obtain analytically, in each case, the first corrections to the linear potential.
\newline\newline
{\textbf{Deep IR.}} This is the simplest case, since we can integrate from $r=0$ to obtain the zero temperature metric,
\begin{equation}
f(r) = \frac{1}{\tilde{\Phi}} \int_{0}^r d\tilde{r}  \, U(\Phi(\tilde{r})) = 2 \pi ^2 \aleph_s \left( r^2 +  \pi \tilde{\Phi} r^3 + \cdots \right) \,,\quad\quad \Phi = \tilde{\Phi} \, r~,
\end{equation}
and we have set $\ell=1$. The $r^2$ term gives the AdS$_2$ metric while the $r^3$ term is the leading correction towards the UV. Going to conformal coordinates we obtain
\begin{equation}
z(r) = \frac{1}{2 \pi ^2 \aleph_s} \left( \frac{1}{r} - \pi  \tilde{\Phi} \log \left(\pi   + \frac{1}{{\tilde{\Phi}} \, r}\right) + \cdots \right) \,,
\end{equation}
which leads to a dilaton and a metric in the conformal gauge of the form
\begin{eqnarray}
\Phi(z) & = & \frac{1}{2 \pi ^2 \aleph_s}\left(\frac{\tilde{\Phi}}{z} -\frac{1}{2 \pi  \aleph_s} \frac{ \tilde{\Phi} ^2 }{z^2} \log \left( \frac{2 \pi^2  \aleph_s z}{\tilde{\Phi}}+ \pi  \right) + \cdots \right) \,,   \label{corr_dil0} \\
e^{2\alpha(z)} & = & \frac{1}{2 \pi ^2 \aleph_s} \left( \frac{1}{z^2} -\frac{ \tilde{\Phi} }{\pi  \aleph_s z^3} \log \left(\frac{2 \pi ^2 \aleph_s z }{\tilde{\Phi}} \right) + \cdots \right) \label{corr_dil1}  \,. 
\end{eqnarray}
We note that $\Phi(\tilde{\Phi}z)$ is independent of $\tilde{\Phi}$, as expected. The presence of a logarithm in the small $z$ expansion (\ref{corr_dil1}) is reminiscent of the Fefferman-Graham expansion in odd dimensional AdS. From the microscopic perspective, it corresponds to the onset of an irrelevant deformation of the deep infrared conformal phase. The logarithm is also present in the near-horizon expansion of extreme Reissner-Nordstr\"om. At finite temperature, the deformation controls the $O(\beta^{-2})$ term in the small temperature expansion of the specific heat. 
\newline\newline
{\textbf{Intermediate IR.}} This case is complicated since perturbatively it is not possible to integrate the potential from $r=0$. However, we expect that the leading correction will be universal and will not depend on the IR details of the theory. The zero temperature metric near the AdS$_2$ boundary becomes
\begin{equation}\label{logmicro}
f(r) = 4 \pi ^2 \left( r-\frac{3}{8\pi \tilde{\Phi} }\right)^2 + \frac{4 \aleph_s }{\pi ^2 \tilde{\Phi} ^2} \log \left(  r -\frac{3}{8\pi \tilde{\Phi} \,} \right) + \cdots \,.
\end{equation}
The first term is the pure AdS$_2$ metric.
 The logarithmic term is the leading correction away from pure AdS$_2$, and it moreover depends on $\aleph_s$. 
There will be further subleading corrections that will depend more sensitively on IR data.
As in the deep IR case, we can use the transformations (\ref{rtoz}) to go to the conformal gauge where
\begin{equation}
z(r) = \frac{1}{ 4 \pi ^2 \left( r-\frac{3}{8\pi \tilde{\Phi} }\right)} -\frac{\aleph_s}{12 \pi ^6 \tilde{\Phi} ^2} \frac{\log \left( r-\frac{3}{8\pi \tilde{\Phi} }\right)}{\left( r-\frac{3}{8\pi \tilde{\Phi} }\right)^3}+ \cdots \,.
\end{equation}
The dilaton and the zero temperature metric in this gauge become,
\begin{eqnarray}
\Phi(z) - \frac{3}{8\pi} & = & \frac{\tilde{\Phi} }{4 \pi ^2 z} - \frac{4 \aleph_s }{3 \pi ^2} \frac{z}{ { \tilde{\Phi}}} \log \, \frac{z}{\tilde{\Phi}} +  \cdots \,, \label{corr_dil2} \\
e^{2\alpha(z)} & = & \frac{1}{4 \pi ^2 z^2} -\frac{20 \aleph_s }{3 \pi ^2 \tilde{\Phi} ^2} \log \frac{z}{ {\tilde{\Phi}} }  + \cdots \,. 
\end{eqnarray}
Note that the correction coming from having a non-vanishing $\aleph_s$ is logarithmic and encodes the effect of a relevant deformation. At finite temperature, the logarithmic term controls the $O(\beta)$ term of the specific heat in a regime where $1/\beta\mathcal{J}$ is much smaller than 1, but greater than $\aleph_s$.
\newline\newline
As a final remark, it may seem unusual that an RG flow in the microscopic theory corresponds to a change in the dilaton potential --  a modification of the bulk theory itself. In AdS/CFT, a relevant deformation is often described by a non-vacuum solution within the same theory. However, this need not be the case in general, particularly if the flow enters regimes where new terms such as higher order corrections or other otherwise suppressed interactions become dominant.\footnote{An extreme case where this could happen is when the deep IR of the RG flow corresponds to a weakly coupled theory in the microscopic dual. If such a situation occurs in a string theoretic bulk, the geometrisation of the RG flow will involve a region where the supergravity approximation breaks down, and is replaced by a description dominated by classical strings.}

%

%

%

\subsection{Bulk dilaton-matter couplings} \label{bulkmattercoupling}

Using the correlation function (\ref{two_point_q}) of the microscopic operator $\zeta$, we now extract properties of the bulk matter couplings.
\newline\newline
\textbf{Case I: $s=0$.} At $s=0$, the microscopic theory is an SYK model with 2 flavours. It contains the near-marginal operator $\zeta$ (\ref{Ozeta}) which has dimension $\Delta_\zeta = 1/q$. At large-$q$, the low-energy two-point function of $\zeta$ is
\begin{equation}\label{2ptszeroA} 
G_\zeta (t_{\text{micro}}) - \frac{1}{4} = -\frac{1}{2q} \left( \log {\cal{J}} |t_{\text{micro}}| + \frac{1}{\mathcal{J} |t_{\text{micro}}|} - \frac{1}{2 (\mathcal{J} |t_{\text{micro}}|)^2} + \ldots   \right) + O\left( q^{-2} \right) \,.
\end{equation}
This follows, for instance, from (\ref{two_point_q}) with $s=0$. Here the coordinate $t_{\text{micro}}$ is dimensionful. The first term, which dominates for $\mathcal{J} |t_{\text{micro}}| \gg 1$, takes the form fixed by conformal invariance $\sim 1/|t_{\text{micro}}|^{2\Delta_\zeta}$ where we further perform a small $\Delta_\zeta$ expansion. In a theory where conformal invariance is preserved, there can be no corrections to this functional form. Thus, the second term in (\ref{2ptszeroA}) must originate from the breaking of conformal invariance. This is in line with the presence of a specific heat linear in temperature, also not allowed in an exactly conformally invariant quantum mechanical theory. The culprit for the breaking, both for the two-point function and the specific heat, is the scale $\mathcal{J}$. To connect to the bulk, we transform (\ref{2ptszeroA}) to frequency space. After incorporating a double-trace deformation as in the discussion leading to (\ref{Q_zeta}), we obtain
\begin{equation} \label{2ptszero}
Q_{\zeta}(\omega_{\text{micro}}) = \frac{1}{\mathcal{J} }\left( -2 \frac{| \omega_{\text{micro}} |}{\mathcal{J}} + \frac{4}{\pi} \frac{| \omega_{\text{micro}} | ^2 \log | \omega_{\text{micro}} |/\mathcal{J}}{\mathcal{J}^2} + \cdots \right)  \,,
\end{equation}
where, for simplicity, we choose a specific normalisation for the two-point function.

From a two-dimensional bulk perspective, the leading piece of the two-point function (\ref{2ptszero}) is a tree-level boundary two-point function of a weakly coupled nearly massless matter field in a background AdS$_2$ geometry with non-trivial dilaton profile (\ref{zeroTDG}). 
We can interpret the deviation from conformality in (\ref{2ptszero}) as a coupling of the matter field to the dilaton field -- i.e. a deviation from a constant $\mathcal{W}(\Phi)$ and/or $\mathcal{V}(\Phi)$ in (\ref{2dmatter}). Since the dilaton field has a non-vanishing profile it is the source of conformal symmetry breaking in the bulk theory. Using the formalism developed in section \ref{sec3}, it is possible to verify that a matter action with $\mathcal{W}(\Phi) = w_0 + w_1 \Phi$ and $\mathcal{V}(\Phi) = 0$ produces the leading correction to the two-point function (\ref{2ptszero}). In terms of bulk field quantities,\footnote{In the massless limit and on-shell, the same correction can be obtained with a constant $\mathcal{W}$  and $\mathcal{V} (\Phi) = v_1 \Phi$. This will produce a term in the matter action of the form $\int du dz \sqrt{g} \Phi \zeta^2$, as in appendix C of \cite{Maldacena:2016upp}. } we find
\begin{equation}
Q_{\zeta}(\omega)  = \ell \tilde{\Phi}^2 \left( - w_0  |\omega| + 2 w_1 \tilde{\Phi} |\omega|^2 \log  |\omega|  + \cdots  \right) ~,
\end{equation}
where the overall normalisation is chosen so that there is no $\tilde{\Phi}$  dependence in $w_0$. Using that $\tilde{\Phi} = 1/ \ell \mathcal{J}$ and $\omega/\ell =  \omega_{\text{micro}}$, we can fix $w_0= 2 $ and $w_1 =2/\pi$.
Finally, $1/q$ corrections to (\ref{2ptszero}) such as the ones computed in \cite{Tarnopolsky:2018env} can provide access to $\mathcal{V}(\Phi)$.
\newline\newline
\textbf{Case II: $s\neq0$.} We would like to use the boundary two-point function for the nearly-marginal operator to assess the couplings of bulk scalars fields. Generally speaking, starting from the microscopic two-point function $Q_\zeta({\omega})$ in (\ref{Q_zeta}) we would like to construct ${\cal{W}}(\Phi)$ and ${\cal{V}}(\Phi)$. For the case at hand, we will make the simplifying assumption that to leading order in the large-$q$ limit, the two-point function $Q_\zeta({\omega})$ can be accounted for entirely through a non-trivial ${\cal{W}}(\Phi)$, such that we can set $\mathcal{V}(\Phi)=0$. This was the case for $s=0$. Recalling (\ref{scalar_eom}), setting $\mathcal{V}(\Phi)=0$ leads to the equation 
\begin{equation} 
 - \partial_z \mathcal{W}(\Phi(z)) \partial_z \zeta_{{\omega}}(z) + \mathcal{W}(\Phi(z))  {\omega}^2 \zeta_{{\omega}}(z) = 0~.
\end{equation}
If $\zeta$ were minimally coupled then ${\cal{W}}(\Phi) = w_0$, leading to $Q({\omega}) = -w_0 |{\omega}|$. Consequently, the expression for $Q_\zeta ({\omega})$ found in (\ref{Q_zeta}) implies $\zeta$ is non-minimally coupled. However both in the UV and the IR, the function $Q_\zeta ({\omega}) \propto -|{\omega}|$ implying that ${\cal{W}}(\Phi(z))$ flows from a constant $w_0$ to another constant value $w_1$ as $z$ flows from the boundary to the deep interior of the bulk. For simplicity, we will fix the normalisation of the two-point function so that $w_0=2$. Then, comparing to (\ref{smallw}) and (\ref{largew}), it turns out that $w_1= 1$. 

As discussed in section \ref{wkb_analysis}, given the boundary two-point function in the large-${\omega}$ limit we can perform a WKB-like analysis to obtain ${\cal{W}}(\Phi(z))$ in a series expansion near the boundary. The first few terms are
\begin{equation} \label{w_wkb}
\mathcal{W}(\Phi(z))  =  2 -\frac{2 \aleph _s}{\pi } \frac{z}{\tilde{\Phi}} +\frac{24 \aleph _s^2}{\pi ^2} \left( \frac{z}{\tilde{\Phi}} \right)^2 + \cdots \,.
\end{equation}
It is straightforward to continue the above expansion to higher orders, as discussed in appendix \ref{wkb_app}. We note that the leading correction is negative, pointing towards the asymptotic value at $z\to \infty$. We can also find the behaviour of $\mathcal{W}(\Phi(z))$ deep in the bulk interior. This is due to the fact that $Q_\zeta (\tilde{\omega})$ is also conformal in the IR. There, we can use (\ref{upsilonR}) to translate the corrections to the conformal propagator in the small-$\tilde{\omega}$ regime --- see (\ref{smallw}) --- to corrections to $\mathcal{W}$ away from the constant $w_1$. We find,
\begin{equation} \label{bulk_interior}
{\cal{W}} (\Phi(z)) = 1 +\frac{1}{2 \pi \aleph_s } \frac{\tilde{\Phi}}{z} + \cdots   \,.
\end{equation}
To the extent that we have tested, the assumption that $\mathcal{V}(\Phi)=0$ to leading order at large-$q$ is indeed valid. 

Expressions (\ref{w_wkb}) and (\ref{bulk_interior}) are sufficient to inform us that for $\aleph_s$ non-vanishing the bulk coupling $\mathcal{W}(\Phi)$ will be non-constant. This should be contrasted with the $\aleph_s=0$ case, where $\mathcal{W}$ was found to be constant to leading order in the large $\mathcal{J}$ expansion. Moreover, given that we know the behaviour of the dilaton in the two regimes --- see (\ref{corr_dil0}) and (\ref{corr_dil2}) --- we find
\begin{eqnarray}
\text{Deep IR:}  & & \mathcal{W}(\Phi) =  1 + \pi \Phi + \cdots   \,, \\
\text{Intermediate IR:}  &  &\mathcal{W}(\Phi) = 2 -\frac{\aleph _s}{2 \pi ^3 \left(\Phi - \frac{3}{8\pi} \right) }  +\frac{\aleph _s^2 \log \left( \Phi - \frac{3}{8\pi}  \right) }{6 \pi ^7 \left( \Phi - \frac{3}{8\pi}  \right) ^3} + \cdots \,.
\end{eqnarray}
We conclude that under the assumption that $\mathcal{V}(\Phi)=0$, we have $\mathcal{W} (\Phi) \neq 16 \pi \Phi$. Comparing to (\ref{minimallycoupled}), we see that this model cannot come from the reduction of a four-dimensional model with minimally coupled matter.

\section{Outlook} \label{sec7}

We have only scratched the surface of what is required to reconstruct a piece of bulk spacetime in two-dimensions. We would like to end with a brief outlook on future directions.
\newline\newline
{\textbf{Landscape of relevant deformations.}} 
The infrared conformal phase of the model (\ref{genSYK}) admits a potentially large class of relevant deformations in the large-$q$ limit. This is because at large-$q$ the weight of the fermion $\Delta_\psi = 1/2q$ becomes small, suggesting that it may be possible to construct a large number of composite operators with $\Delta < 1$. A natural generalisation of the deformation studied here is 
\begin{equation}
H_{\text{gen}}(\lambda_i) = H_q + \sum_{i=1}^n  \lambda_i H_{q_i}~,
\end{equation}
with all the $q_i<q$ scaling with $q$ at large $q$. It would be interesting to understand if $H_{\text{gen}}(\lambda_i)$ is still under computational control. It may also be interesting to consider models with various $H_q$ operators whose $q$ values differ by an $\mathcal{O}(1)$ amount. Can we construct a broader class of controllable RG flows in these models and assess their holographic features in along the flow and in the deep infrared?
\newline\newline
{\textbf{Minimal coupling and fine tuning.}}
From a higher-dimensional perspective of dilaton-gravity models in two-dimensions, the dimensionally reduced matter fields couple to the dilaton-gravity sector in a particular way. For example, four-dimensional minimally coupled matter manifests itself in the two-dimensional dilaton-matter couplings (\ref{minimallycoupled}).  The reduction of a four-dimensional model with non-minimal couplings would differ. Can we microscopically engineer quantum mechanical models dual to two-dimensional models whose quadratic action mimics an action stemming from the dimensional reduction of minimal coupled matter? It would be interesting to understand the degree of fine-tuning required to mimic four-dimensional physics. 

Perhaps relatedly, we have discussed various energy conditions from a macroscopic perspective and their ensuing constraints. What property (if any) of quantum mechanical microscopic models, at least in the large $N$ limit and for a particular class of states, leads to the null-energy or some other energy condition in the bulk dual \cite{Parikh:2015ret}.
\newline\newline
{\textbf{A geometrically regularised soft mode.}}
We have argued that for $s \ll 1$ the flow geometry dual to the microscopic model (\ref{Hdeformed}) can be viewed as interpolating between two (near) AdS$_2$ geometries. Thermodynamically, we saw two different temperature regimes: one for intermediate temperatures and the other in the deep IR. In both regimes, the specific heat was approximately linear. This suggests that there is an effective soft mode (perhaps of the Schwarzian type) residing at the boundary of the AdS$_2$ in the deep IR. If so, the deep IR soft mode is regularised within the strongly coupled regime of SYK. Indeed, from the effective action (\ref{eff_act}) point of view the approximate reparametrisation symmetry of the deep IR, governed by the $G_{\psi\psi}^{q/2} G_{\xi\xi}^{q/2}$ term, is already broken by the $G_{\psi\psi}^q G_{\xi\xi}^q$ term. 

It would be interesting to better understand the UV regularisation of the deep IR soft mode and the geometric manifestation thereof. Relatedly, as a simple test of the holographic hypothesis that the full RG flow is captured by a dilaton-gravity model it would be interesting to compute the bulk four-point function and compare to the one predicted by the microscopic theory \cite{Jiang:2019pam}. 
\newline\newline
{\textbf{Toward a microscopic realisation of the centaur geometry?}}
The presence of the coupling $s$ in (\ref{Hdeformed}) gives us a new dimensionless parameter, in addition to $N$ and $q$, in the space of SYK-like models. Unlike $N$ and $q$, the parameter $s$ is not an integer in the microscopic theory. Although unitarity imposes $s\in\mathbb{R}$, we can consider analytically extending $s$ to the complex plane. This may give rise to new behaviour in the deep infrared of the flow. This is briefly explored in appendix \ref{F_app}. One  possibility of interest is a specific heat that is positive and depends linearly on the temperature for an intermediate temperature range but flows to a negative linear dependence on the temperature in the deep infrared.  Such a flow would be captured by a dilaton-gravity theory whose dilaton potential goes as $+\Phi$ for some region before changing to a $-\Phi$ dependence. This behaviour occurs for (\ref{potential_deep}) and (\ref{potential_int}) if one (na\"ively) continues $\aleph_s \to - \aleph_s$. Such theories were macroscopically explored in \cite{Anninos:2017hhn,Anninos:2018svg}, and admit asymptotically AdS$_2$ geometries that flow to a portion of dS$_2$ in the deep interior. Can we make sense of theories (\ref{Hdeformed}) with $s\in\mathbb{C}$, perhaps along the lines of \cite{Gorbenko:2018ncu}?
%

\section*{Acknowledgements}

We would like to gratefully acknowledge Leonel Queimada for collaboration at the initial stages of the project. We also acknowledge many insightful discussions with B. Czech, A. Levine, S. Murthy, J. Pedraza, V. Rosenhaus, R. Russo, J. Turiaci, D. Vegh, and especially T. Anous, D. Hofman and E. Shaghoulian. Our research is funded by the Royal Society under the grants ``The Atoms of a deSitter Universe" and ``The Resonances of a deSitter Universe". The work D.A.G. is further funded by the ERC Consolidator Grant N. 681908, ``Quantum black holes: A microscopic window into the microstructure of gravity".

\appendix

\section{More on constraints from the null energy condition} \label{nec}
Here we will slightly generalise the metric (\ref{ineq}) to allow for more general transverse spaces. Assume that the four-dimensional metric is given by
\begin{equation}
ds^2 = -f(r) dt^2 + \frac{dr^2}{f(r)} + g(r)^2 ds_k^2 \,,
\end{equation}
where $ds_k^2$ is either the metric of the 2-dimensional sphere, plane or hyperbolic space. Namely, for 
\begin{eqnarray}
k = +1: & \,\,\, d\Omega_2^2 \,, \\
k  =  0 :   & \,\,\, (dx^2 + dy^2) \,, \\
k  =  -1:  & \,\,\, d\Sigma_2^2 \,.
\end{eqnarray}
The case $k=1$ is discussed in the main text. By imposing the null energy condition, we now obtain the following inequalities,
\begin{equation}
-f g g'' \ge 0~, \quad\quad  2k +g^2 {f''} \ge  {f}  \, (g^2)''~.
\end{equation}
The first one is the same for all $k$. It is interesting to analyse the different constraints the second one gives. As an example, take $g(r)=a_0 + a_1 r$. Then
\begin{eqnarray}
k + \frac{g^2}{2} f'' \geq a_1^2 f \,.
\end{eqnarray}
Given the above inequality, we see that only positive values of $k$ admit solutions that interpolate between $f''>0$ and $f''<0$. As in the main text, it is useful to define a new variable, $F(r) \equiv f(r)/g(r)^2$. In this new variable, the second inequality becomes,
\begin{eqnarray}
2k + (g^4 F')' \geq 0 \,.
\end{eqnarray}
Picking $r$ to be an monotonically increasing coordinate originating at $r_h$, where $f(r_h) = 0$ and $g(r_h)>0$, we can integrate the above expression to get,
\begin{eqnarray}
2 k (r-r_h) + g(r)^4 F'(r) - g(r_h)^4 F'(r_h)   \geq 0 \,.
\end{eqnarray}
Assuming that the last term is either vanishing or positive, we observe that only positive $k$ allows for a change in the sign of $F'$.

A similar analysis with a metric of the form $ds^2 = dr^2 + e^{2A(r)} (-dt^2 + d\vec{x}^2)$ leads to the so-called $c$-theorems in holography \cite{Myers:2010xs}.

\section{Two-point function of the centaur geometry} \label{app_centaur}

\begin{figure}[h!]
        \centering
        \subfigure[Boundary two-point function]{
                \includegraphics[scale=0.58]{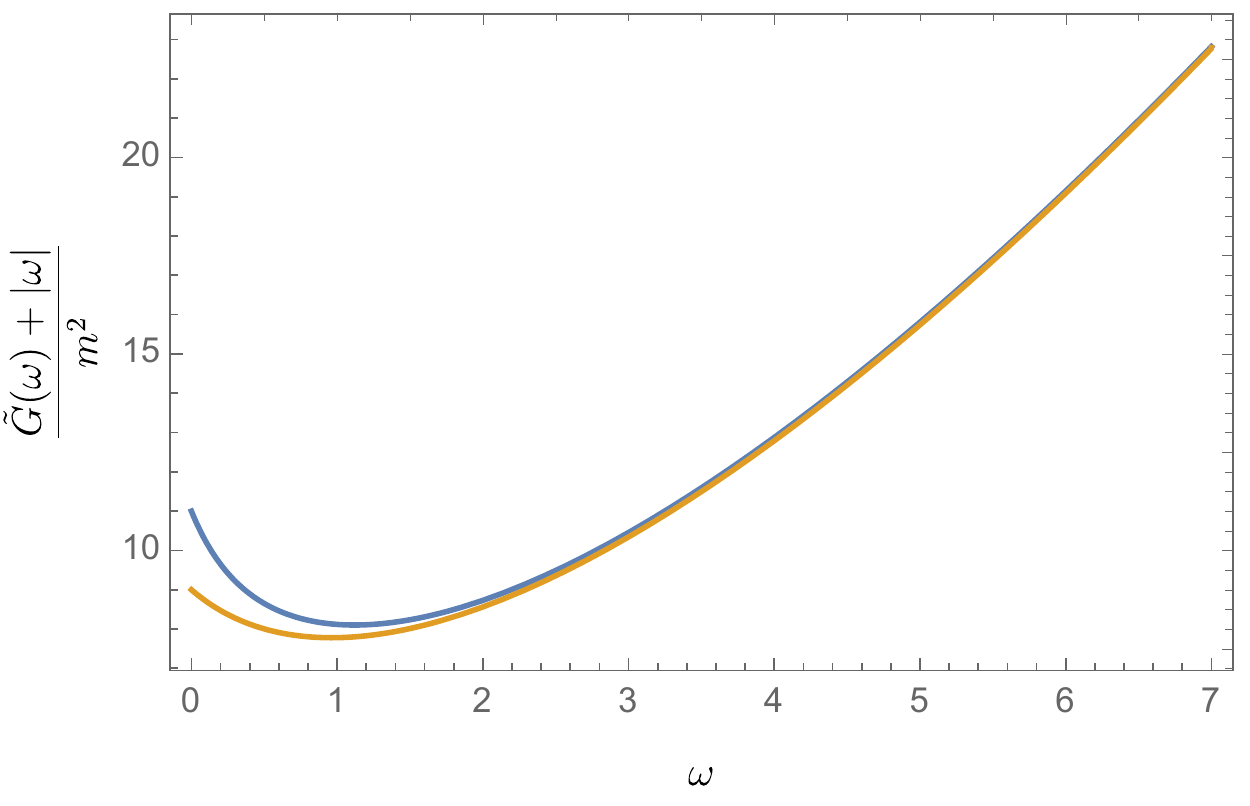} \label{fig_centaur_G}} \quad\quad
         \subfigure[Bulk metric]{
                \includegraphics[scale=0.58]{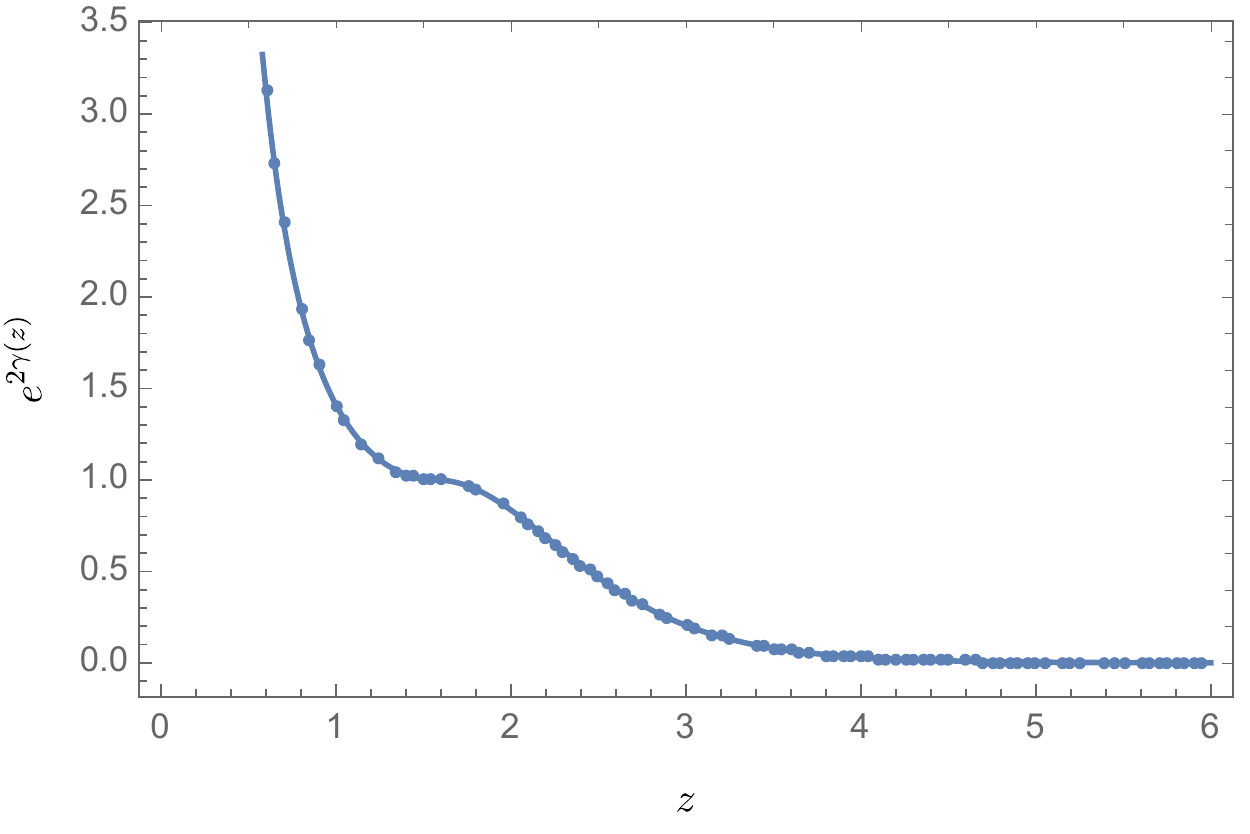} \label{fig_centaur_metric}}
                 \caption{{\footnotesize (a) Correction to the boundary two-point function for the centaur geometry in blue and the AdS$_2$ black hole in yellow ($e^{2\gamma(z)} = \sinh^{-2} z$). We see both curves rapidly overlapping as $\omega$ grows larger, while differences appear for small-$\omega$. $z_c=0.1$ in the plot. \newline
                 (b) The metric of the centaur geometry as a function of bulk coordinate $z$. The blue dots are the numerical metric reconstruction from the boundary two-point function in (\ref{centaur_G}), while the solid curve is the exact plot of the metric in (\ref{centaur_metric}).  }}
\end{figure}

An interesting example of a flow geometry is the centaur geometry \cite{Anninos:2017hhn,Anninos:2018svg}. This geometry interpolates between an AdS$_2$ patch near the boundary and a part of the static patch of dS$_2$ in the interior. The simplest centaur geometry is given by
\begin{eqnarray} \label{centaur_metric}
e^{2\gamma(z)} =   \begin{cases}
     \sin^{-2}  z & \text{for} \,\,\, 0<z<\pi/2 \,,\\
     \cosh^{-2} \left(z- \frac{\pi}{2} \right) &\text{for} \,\,\, \pi/2<z \,.
    \end{cases}       
\end{eqnarray}
From (\ref{greenbdy}) we obtain the corresponding boundary two-point function for a nearly massless scalar:
\begin{eqnarray} \label{centaur_G}
\tilde{G}(\omega) & = & - |\omega| - m^2 \left( -i+\frac{1}{z_c}+2 \gamma  \left| \omega \right| -i \pi  \left| \omega \right| +\left| \omega \right|  \log \left(4 z_c^2\right) + 2 \left| \omega \right|  \psi ^{(0)}(i \left| \omega \right| ) - \right.  \\
& & \left. -e^{-\pi  \left| \omega \right| } \left(\left| \omega \right|  H_{\frac{\left| \omega \right| }{2}}-\left| \omega \right|  H_{\frac{\left| \omega \right| -1}{2}}+\left| \omega \right|  \psi ^{(0)}\left(\frac{i \left| \omega \right| }{2}\right)-\left| \omega \right|  \psi ^{(0)}\left(\frac{i \left| \omega \right| }{2}+\frac{1}{2}\right)+(-1-i)\right) \right) \,, \nonumber
 \end{eqnarray}
where $H_n$ is the $n^{\text{th}}$ harmonic number. We display (\ref{centaur_G}) in figure \ref{fig_centaur_G} and compare it with the boundary two-point function of the AdS$_2$ black hole. Interestingly, for large $\omega$,
\begin{eqnarray}
\tilde{G}(\omega) & \approx &  - |\omega| -  m^2\left( \frac{1}{z_c} +2 \left| \omega \right|  \log (2 e^{\gamma_E} |\omega| z_c) +\frac{1}{6 \left| \omega \right| } +\frac{1}{60 \left| \omega \right| ^3}+\frac{1}{126 \left| \omega \right| ^5} + O (|\omega|)^{-7}   \right. \nonumber \\
 &  & -  \left. e^{-\pi  \left| \omega \right| } \left(\frac{1}{2 \left| \omega \right| ^3} + \frac{17}{4 \left| \omega \right| ^7} + O(|\omega|)^{-11} \right) \right) \,.
\end{eqnarray}

It is also possible to use (\ref{inverseLap}) to numerically invert the two-point function and obtain back the centaur metric, as shown in figure \ref{fig_centaur_metric}.

\section{Details of the WKB expansion} \label{wkb_app}

In this appendix, we give some details on the calculation of the function ${\mathcal{W}}(\Phi(z))$ using the WKB approximation. Our starting point is the Euclidean action
\begin{equation}
S_E  =  \frac{1}{2} \int dt dz \, \mathcal{W}(\Phi(z))\left( (\partial_z \zeta(t,z))^2 +  (\partial_t \zeta(t,z))^2 \right)~.
\end{equation}
We assume that $\mathcal{W}(\Phi(z))$ is purely a function of $\Phi(z)$, and its dependence on $\tilde{\Phi}$ is fixed entirely through its functional dependence on $\Phi$. We set $\ell=1$ in what follows. On-shell, the action is given by
\begin{equation}
S_{\text{on-shell}} = - \frac{ \mathcal{W}_0}{2} \int_{\mathbb{R}} \frac{d\omega}{2\pi} \xi_\omega   \partial_z \zeta_{-\omega}(z)|_{z=0}~, \quad\quad \mathcal{W}_0 \equiv \mathcal{W}(\Phi(0))~.
\end{equation}
We want to solve the equation of motion at large-$\omega$,
\begin{equation} 
 - \partial_z \mathcal{W}(\Phi(z)) \partial_z \zeta_{\omega}(z)+ \mathcal{W}(\Phi(z))  \omega^2 \zeta_\omega(z) = 0~.
\end{equation}
It follows from the general structure of the dilaton-gravity solution that $\Phi(\tilde{\Phi} z) = \Phi(z)$, as was observed in (\ref{corr_dil0}). As such, it is natural to consider the problem in terms of the variables $\tilde{z} = z/\tilde{\Phi}$ and $\tilde{\omega} = \tilde{\Phi} \omega$. Our equation of motion becomes
\begin{equation} 
 - \partial_{\tilde{z}} \mathcal{W}(\Phi(\tilde{z})) \partial_{\tilde{z}} \zeta_{\tilde{\omega}}(\tilde{z})+ \mathcal{W}(\Phi(\tilde{z}))  \tilde{\omega}^2 \zeta_{\tilde{\omega}}(\tilde{z}) = 0~.
\end{equation}
The dependence on $\tilde{\Phi}$ can be reinstated by rescaling $\tilde{z}$ and $\tilde{\omega}$. In order to put the above equation in the Schr\"odinger form, we introduce a coordinate $u$
\begin{equation}
\tilde{z} = \int_0^u {\mathcal{W}} (\tilde{z}(u')) du' \,,
\end{equation}
leading to
\begin{equation} \label{WKB_equation}
-\partial_u^2  \zeta_{\tilde{\omega}}(u) + \tilde{\omega}^2 {\mathcal{W}}^2 (\tilde{z}(u) ) \zeta_{\tilde{\omega}}(u) =0 \,.
\end{equation}
Employing the standard WKB ansatz 
\begin{equation}\label{WKBapp}
\zeta_{\tilde{\omega}} (u) = \tilde{\xi}_{\tilde{\omega}} \, \exp \left( \frac{1}{\delta} \sum_{n=0}^\infty \delta^n Q_n(\tilde{\omega},u) \right) \,,
\end{equation}
where the normalisation $\tilde{\xi}_{\tilde{\omega}}$ is chosen such that the boundary value of $\zeta_\omega(0)$ is $\xi_\omega$. We can now solve (\ref{WKB_equation}) in powers of $\delta$
\begin{equation}
\frac{1}{\delta} \left( \sum_{n=0}^\infty \delta^n Q''(\tilde{\omega},u) \right) + \frac{1}{\delta^2} \left( \sum_{n=0}^\infty \delta^n Q'(\tilde{\omega},u) \right)^2 = \tilde{\omega}^2 {\cal{U}}(u)^2 \,,
\end{equation}
where primes denote derivatives with respect to $u$ and $\mathcal{U}(u) \equiv \mathcal{W}(\tilde{z}(u))$. To leading order we have
\begin{equation}
 Q_0 (\tilde{\omega},u) = -|\tilde{\omega}| \tilde{z}(u) \,.
 \end{equation}
The next orders yield
 \begin{eqnarray}
 Q_0'' + 2 Q_0' Q_1'  =   0 & \to & Q_1 = -\frac{1}{2} \log \tilde{z}'(u) = -\frac{1}{2} \log {\cal{U}} \,, \\
 Q_1'' + Q_1'^2 + 2 Q_0' Q_2' = 0 & \rightarrow & Q_2' = - \frac{Q_1'' + Q_1'^2}{2Q_0'} \,, \\
 Q_2'' + 2 Q_1' Q_2' + 2 Q_0' Q_3'  =  0 & \rightarrow & Q_3'= \frac{-2 Q_1' Q_2' - Q_2''}{2 Q_0'} \,,
\end{eqnarray}
where each term depends on the solution to the previous ones. From these terms, we obtain the following macroscopic two point function,
\begin{equation}
Q_{\zeta} (\tilde{\omega})  = \mathcal{U}_0  \left( -|\tilde{\omega}| - \frac{1}{2} \frac{\mathcal{U}'(0)}{\mathcal{U}(0)^2} + \frac{3 \mathcal{U}'(0)^2-2 \mathcal{U}(0) \mathcal{U}''(0)}{8 \mathcal{U}(0)^4 |\tilde{\omega}|} + \ldots \right) \,. \label{wkb_W}
\end{equation}
We can compare to the microscopic model, order by order in inverse powers of $\tilde{\omega}$. We recall here from (\ref{largew}) that the two-point function for our particular microscopic model is given by
\begin{equation}
Q_\zeta (\tilde{\omega})  = 2  \left( -  \left| \tilde{\omega} \right|+\frac{2 \aleph_s}{\pi }-\frac{4}{\pi ^2} \frac{\aleph_s^2}{\left| \tilde{\omega} \right|} + \cdots \right) \,,
\end{equation}
where we fixed the overall normalisation of the two-point function, and we have used the identification $\tilde{\Phi} = 1/\mathcal{J}$. Comparing the above expressions we find that 
\begin{equation}
\mathcal{U} (u) =  2 - \frac{16 \aleph _s}{\pi } u + \frac{160 \aleph _s^2}{\pi ^2} u^2 + \cdots \,.
\end{equation}
In order to reconstruct $\mathcal{W}$, we go back to $\tilde{z}$ coordinates. Close to the boundary,
\begin{eqnarray}\label{app_w_z}
\mathcal{W}(\Phi(\tilde{z}\approx0))  & = &   \mathcal{U}(0) \left(1 + \frac{ \mathcal{U}'(0)}{ \mathcal{U}(0)^2} \tilde{z} + \frac{ \mathcal{U}(0) \mathcal{U}''(0)  - \mathcal{U}'(0)^2}{2\mathcal{U}(0)^4} \tilde{z}^2 + \ldots \right) \,,
\end{eqnarray}
where we recall that the primes indicate $u$-derivatives. For our model this becomes,
\begin{equation}
\mathcal{W}(\Phi(\tilde{z}\approx0))  =  2 -\frac{2 \aleph _s}{\pi } \tilde{z} +\frac{24 \aleph _s^2}{\pi ^2} \tilde{z}^2 + \cdots   \,,
\end{equation}
that is the final expression shown in (\ref{w_wkb}) and can be systematically improved to the desired order.

\section{The thermal partition function of the microscopic model}\label{F_app}

In this appendix we give some details on the derivation of the effective action and analyse the free energy of the microscopic model both analytically and numerically for different values of the coupling $s^2$. 

\subsection{Derivation of the effective action}
We start with the derivation of the thermal partition function for the microscopic model under consideration, (\ref{Hdeformed}). This follows closely the standard techniques used in the SYK literature, so we will just point out the basic results. The action is given by
\begin{eqnarray}
S_{\text{micro}} & = & \int d\tau \left( \frac{1}{2} \sum_{i=1}^{N/2} (\psi_i \partial_\tau \psi_i +  \xi_i \partial_\tau \xi_i) + \frac{i^{q}}{(q!)^2} \sum_{I_1} J_{I_1}^{(q)} (\psi_{i_1} \cdots \psi_{i_{q}})   (\xi_{j_1} \cdots \xi_{j_{q}}) + \right. \nonumber \\
& & \hspace{4cm} \left.   + s  \frac{i^{q/2}}{((q/2)!)^2} \sum_{I_2} J_{I_2}^{(q/2)} (\psi_{i_1} \cdots \psi_{i_{q/2}})   (\xi_{j_1} \cdots \xi_{j_{q/2}})    \right) \,, 
\end{eqnarray}
where the collective indexes are $I_1 = i_1, \cdots, i_q; j_1, \cdots, j_q$ and $I_2 = i_1, \cdots, i_{q/2}; j_1, \cdots, j_{q/2}$ and each $i$ and $j$ runs from 1 to $N/2$.
We want to compute the partition function,
\begin{equation}
Z = \int dJ^{(q)}_{I_1} dJ^{(q/2)}_{I_2} P[J^{(q)}_{I_1}] P[J^{(q/2)}_{I_2}] \int D\psi_i D\xi_i  e^{-S_{\text{micro}}} \,,
\end{equation}
where the probability distributions $P[J]$ for the random couplings are both Gaussian with zero mean and variance given by (\ref{gaussian}). We can integrate out the couplings and define bi-local variables,
\begin{equation} \label{G_flavour}
G (\tau_1, \tau_2) = \frac{2}{N} \sum_{i=1}^{N/2} \begin{pmatrix} \psi_i (\tau_1) \psi_i(\tau_2) && \psi_i (\tau_1) \xi_i(\tau_2) \\
\xi_i (\tau_1) \psi_i(\tau_2) && \xi_i (\tau_1) \xi_i(\tau_2) \end{pmatrix} \,.
\end{equation}
In this model, $G_{ab}$ becomes a matrix with flavour indexes. To introduce $G_{ab}$ in the partition function, we also introduce $\Sigma_{ab}$ variables as Lagrange multipliers to enforce the constraint (\ref{G_flavour}). Integrating out the fermions we obtain the following effective action,\footnote{It is useful to note that the determinant for a block diagonal matrix is given by $\det \begin{pmatrix} A & B\\ C & D \end{pmatrix} = \det (A - B D^{-1} C) \det D $.}
\begin{eqnarray} \label{eff_act}
\frac{S_{\text{eff}}[G_{ab}, \Sigma_{ab}]}{N} & = & -\frac{1}{4} \text{tr}\log (\partial_\tau - \Sigma_{\xi \xi}) -\frac{1}{4} \text{tr}\log \left( (\partial_\tau - \Sigma_{\psi\psi}) - \Sigma_{\psi \xi} \left( \partial_\tau - \Sigma_{\xi \xi} \right)^{-1} \Sigma_{\xi \psi} \right) + \nonumber \\
& & +  \int d\tau_1 d\tau_2 \left(  \sum_{\substack{a=\psi,\xi \\ b=\psi,\xi}} \frac{1}{4} \Sigma_{ab} G_{ab} - \frac{J_q^2}{2 q^2} G_{\psi\psi}^q G_{\xi\xi}^q - s^2 \frac{J_{q/2}^2}{2 (q/2)^2} G_{\psi\psi}^{q/2} G_{\xi\xi}^{q/2} \right) \,.
\end{eqnarray}
By symmetry, $G_{\psi\psi} = G_{\xi\xi} \equiv G_\psi$, $\Sigma_{\psi\psi} = \Sigma_{\xi\xi} \equiv \Sigma_\psi$, $G_{\psi\xi} =- G_{\xi \psi}$  and $\Sigma_{\psi\xi} =-\Sigma_{\xi \psi}$. Varying this action with respect to $G_\psi$ and $\Sigma_\psi$, we obtain the Schwinger-Dyson equations in (\ref{SD1})-(\ref{SD2}). Moreover we see that, at leading order, $G_{\psi \xi}=\Sigma_{\psi\xi} = 0$. Expanding the action for fluctuations around this saddle, we confirm that the two-point function for the scalar $\zeta$ in the large $N$ limit is given by $G_\zeta = G_\psi^2/N$, as stated in (\ref{G_scalar}).

\subsection{Computing the free energy from the two-point function}
In this subsection, we will be interested in the thermodynamics of the system at leading order in the large-$q$ expansion. By computing the action (\ref{eff_act}) on-shell, using parameterisation (\ref{Gfermion}), it is possible to show that the free energy $F$ at large-$q$ is given by \cite{Jiang:2019pam}
 \begin{equation} \label{free_energy}
 S_{\text{eff}}^{\text{on-shell}} \equiv \beta F = - \frac{N \beta}{8 q^2} \int_0^\beta d\tau \left[ \frac{1}{2} (\partial_\tau g(\tau))^2 +  \mathcal{J}^2 \left(2 s^2 e^{g(\tau)} + \frac{1}{2} e^{2g(\tau)} \right) \right] \,,
 \end{equation}
 where $\mathcal{J}$ is defined below (\ref{g_eq}) in the main text. Note that written in this way, the only input needed to compute the free energy is the thermal two-point function $g(\tau)$, that is given by (\ref{g_finite_T}) and that we recall here,
\begin{equation}\label{two_app}
e^{g(\tau)} = \frac{2 \nu ^2}{\sqrt{(\beta {\cal{J}})^2 \nu ^2+ s^4 (\beta  {\cal{J}})^4} \cos (\nu  (\frac{2 \tau}{\beta}-1))+ s^2 (\beta {\cal{J}})^2}~,
\end{equation}
with 
\begin{equation}
\cos \nu =\frac{2 \nu ^2- s^2 (\beta {\cal{J}})^2}{\sqrt{(\beta {\cal{J}})^2 \nu ^2+ s^4 (\beta {\cal{J}})^4}}~.
\end{equation}
So, in principle given $\nu$ and $s$, we can first compute $g(\tau)$ and $\beta$ and from there, integrate to get the thermodynamic free energy of the system. In this section, we will be interested only in the deep infrared behaviour, so $\beta \J \gg \text{max} (1,s^{-2}))$. At large $\beta \mathcal{J}$, we expect the free energy to behave as
\begin{equation} \label{free_low_temps}
\beta F = -\frac{N}{q^2} \left( f_0 \beta \J + f_1 + \frac{f_2}{\beta \mathcal{J}} + \cdots \right) \,,
\end{equation}
with $f_i$ numbers that might depend on $s$. In the main text, we were mostly concerned with the entropy $S$ and the specific heat $C$, that can be obtained from the free energy by
\begin{equation}
S = -(1-\beta \partial_\beta) (\beta F)~, \quad\quad C = -\beta^2 \partial^2_\beta (\beta F) \,.
\end{equation}
Note that $f_0$ contributes neither to the entropy nor to the specific heat. The correction to the zero-temperature entropy is given by $f_1$ while $2f_2$ is the coefficient of the linear-in-temperature specific heat. In the following, we will compute these quantities for different values of $s$.

\subsubsection{$s^2=0$}
The simplest case is to turn off the coupling and we expect the thermodynamics to be that one of the standard SYK model with a $2q$ interaction. We should recover (\ref{series_SYK_deepIR}) with $q \to 2q$. At $s^2=0$, (\ref{two_app}) becomes
\begin{eqnarray}
e^{g(\tau)} = \frac{2 \nu }{\beta \J} \sec \left(\nu  \left(2 \tau/\beta -1\right)\right) \quad\quad \text{with}  \quad\quad  \cos \nu =\frac{2 \nu }{\beta \J } \,.
\end{eqnarray}
It is clear that large $\beta \J$ means $\nu \sim \pi/2$, and in fact expanding the equation on the right we get,
\begin{equation}
\nu = \frac{\pi }{2}-\frac{\pi }{\beta \J}+\frac{2 \pi }{(\beta \J)^2}-\frac{\pi  \left(24+\pi ^2\right)}{6 (\beta \J) ^3}+O\left(\beta \J\right)^{-4} \,.
\end{equation}
Note it is important to get to this order $(\beta \J)^{-3}$ in the expansion, to get the right coefficient of the specific heat. In this simple case, one can integrate (\ref{free_energy}) analytically and expand for large $\beta \J$ to obtain,
\begin{equation}
\beta F = \frac{N}{q^2} \left( -\frac{\beta \J}{4}+\frac{\pi ^2}{16}-\frac{\pi ^2}{8 \beta \J}+ \cdots \right) \,,
\end{equation}
which is the expected result for the $2q$-SYK model.

\subsubsection{$s^2>0$}
\begin{figure}[h!]
        \centering
        \subfigure[Free energy for $s^2=0.01$]{
                \includegraphics[scale=0.6]{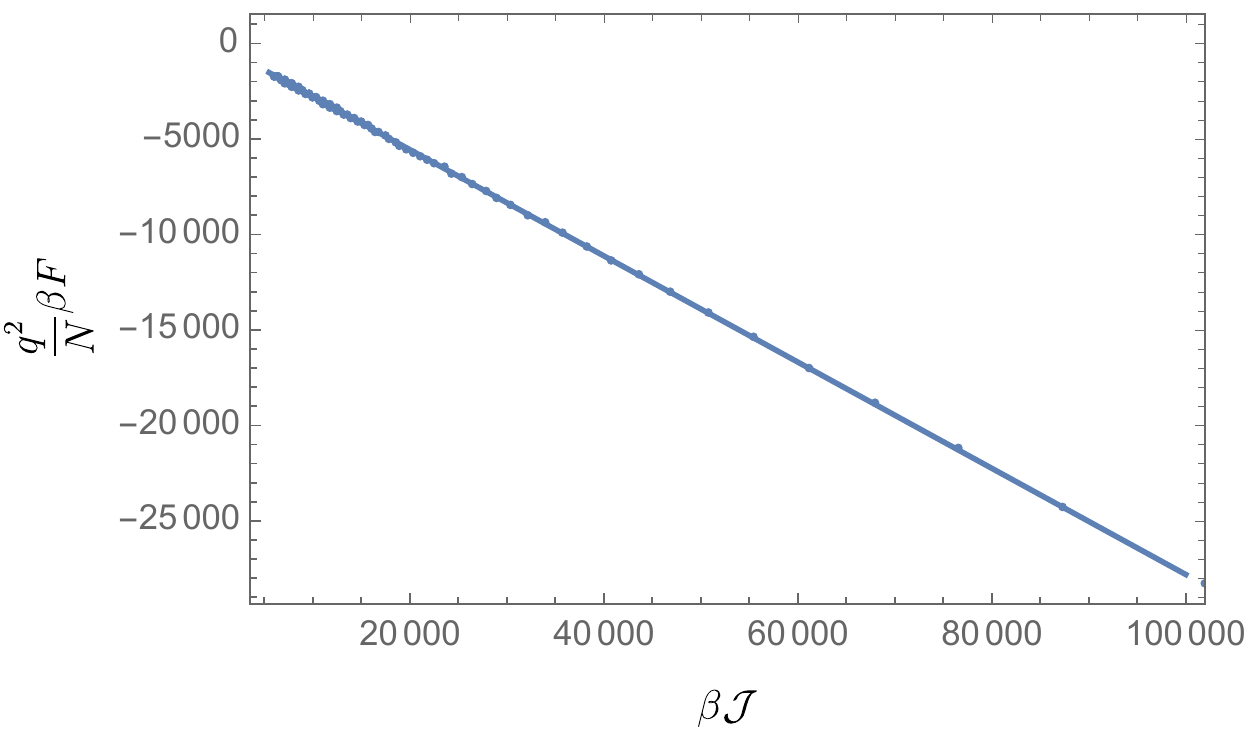} \label{fig_F_positive}} \quad\quad
         \subfigure[Specific heat]{
                \includegraphics[scale=0.56]{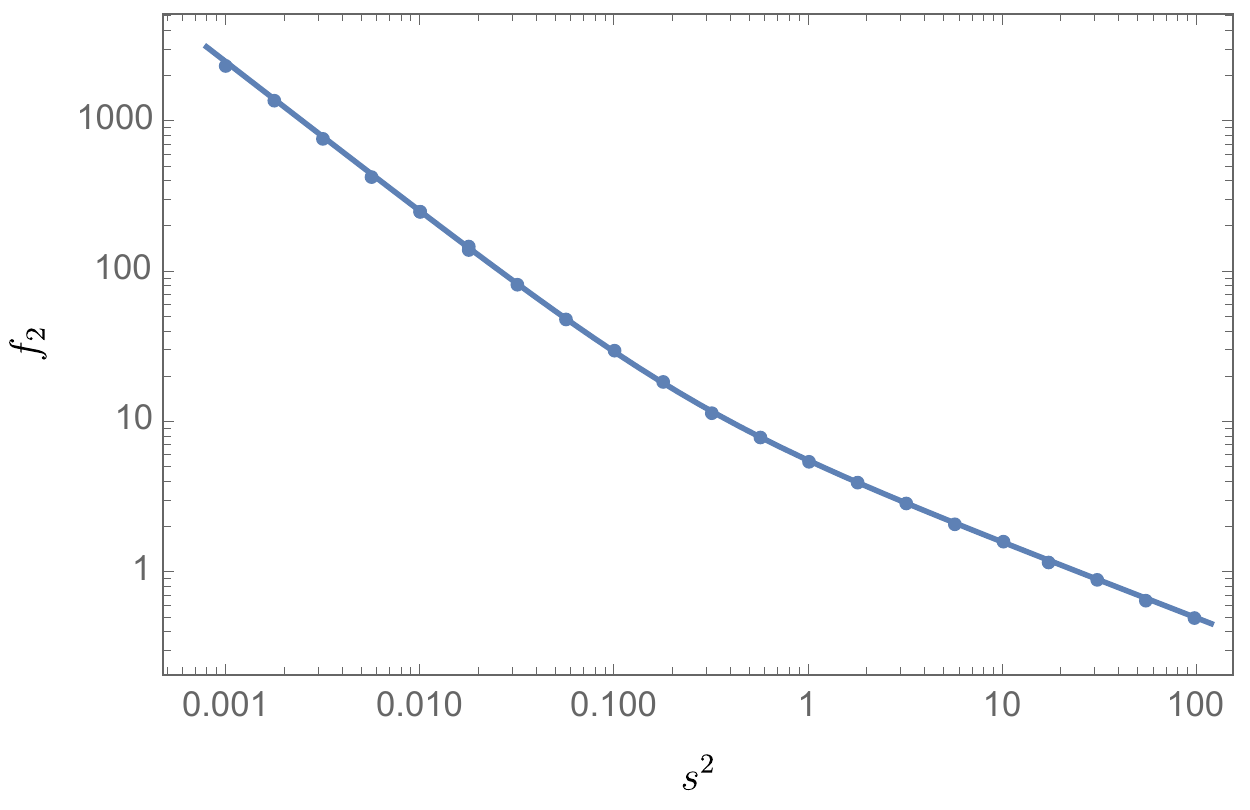} \label{fig_C_positive}}
                 \caption{{\footnotesize (a) The free energy as a function of the inverse temperature for $s^2=0.01$. The dots correspond to numerically integrating (\ref{two_app}) for different temperatures, while the solid line corresponds to the analytic result at this value of the coupling. \newline
                 (b) The coefficient $f_2$ as a function of $s^2$ in a logarithmic scale. The dots correspond to numerically integrating the free energy at each $s^2$ and fitting by a function of the form of (\ref{free_low_temps}). The solid line is the analytic result $f_2 = \pi^2 \sqrt{4s^2+1} /4s^2$. }}
\label{fig_s_positive}
\end{figure}

Now we study the deformed model. The first difference with respect to the $s^2=0$ case comes when expanding $\nu$ for large $\beta \J$, using the expression in the right side of (\ref{two_app}). In this case, $\nu$ becomes close to $\pi$,
\begin{equation}
\nu = \pi -\frac{\pi  \sqrt{1+ 4 s^2}}{s^2 \beta \J}+\frac{4 \pi  s^2+\pi }{s^4} \frac{1}{(\beta \J)^2}+\frac{\pi ^3 (8 s^2+3)-8 \pi  (1+ 4 s^2)^2}{8 s^6 \sqrt{1+ 4 s^2}} \frac{1}{(\beta \J) ^3}+O\left(\beta \J\right)^{-4} \,.
\end{equation}
Following the approach of \cite{Jiang:2019pam} which leads to (\ref{temperature}) and (\ref{Zexact}), we find the small temperature expansion
\begin{equation} \label{free_s_positive}
 \beta F = \frac{N}{q^2} \left( -f_0 \beta \J + \frac{\pi ^2}{4}-\frac{\pi ^2 \sqrt{1+ 4 s^2}}{4 s^2} \frac{1}{\beta \J} +  \cdots \right) \,,
 \end{equation}
with
\begin{equation}
f_0 = \frac{1}{4} \left(\sqrt{1+4 s^2}+2 s^2 \coth ^{-1}\left(\frac{1+2 s^2}{\sqrt{1+4 s^2}}\right)\right) = \frac{1}{4} + \frac{1}{2} s^2 \left(1-\log s^2\right) + O(s^4) \,.
\end{equation}
Comparing with the free energy in the zero-coupling case, we already see that the limit of $s^2 \to 0$ will be discontinuous. This is expected since the deep IR, whenever the coupling is turned on will be dominated by a $q$-SYK model while exactly at $s^2=0$, the thermodynamics will be those of a $2q$-SYK, as shown in the previous subsection. We can check this expression by integrating (\ref{two_app}) numerically. This is shown in figure \ref{fig_s_positive}, where we first plot the free energy for a specific value of $s^2=0.01$ and then we plot the coefficient of the linear-in-temperature term as a function of $s^2$. As shown, both agree to a high degree of accuracy. Note that the specific heat is always positive, as expected.

Finally, it is interesting to note that we can add a small complex part to the coupling, $s^2= s^2 + i \delta$, while keeping the temperature real. This will produce a complex free energy but interestingly, the form (\ref{free_s_positive}) will remain unchanged (with complex $s^2$ now). This can be checked numerically for a variety of small $\delta$, obtaining analogue results to those in figure \ref{fig_s_positive} and showing that it seems reasonable to take the limit of $\delta \to 0$ in this case.

\subsubsection{$s^2<0$}
Continuing $s^2$ to negative values is, of course, more subtle. We will restrict ourselves to $-1/4<s^2<0$, so that $\sqrt{1+4s^2}$ appearing in, for instance, (\ref{free_s_positive}) stays real. The first thing to note is that when $s^2<0$, at large $\beta \J$, $\nu$ doesn't approach $\pi$ as in the positive case, but $2\pi$. In fact, we obtain,
\begin{equation}
\nu = 2 \pi -\frac{2 \pi  \sqrt{1+ 4 s^2}}{  s^2 \beta \J}+\frac{2 \pi  (1+ 4 s^2)}{s^4} \frac{1}{(\beta \J)^2}+\frac{\pi ^3 (8 s^2+3)-2 \pi  (1+ 4 s^2)^2}{s^6 \sqrt{1+ 4 s^2}} \frac{1}{(\beta \J)^3}+ \cdots \,.
\end{equation}

The second important observation is that when $s^2<0$ at leading order in the large $N$ and large-$q$ limit, the two point function becomes discontinuous at certain times. At large $\beta \J$, this happens close to $\tau \approx 0, \beta/2, \beta$.

\begin{figure}[h!]
        \centering
        \subfigure[]{
                \includegraphics[scale=0.4]{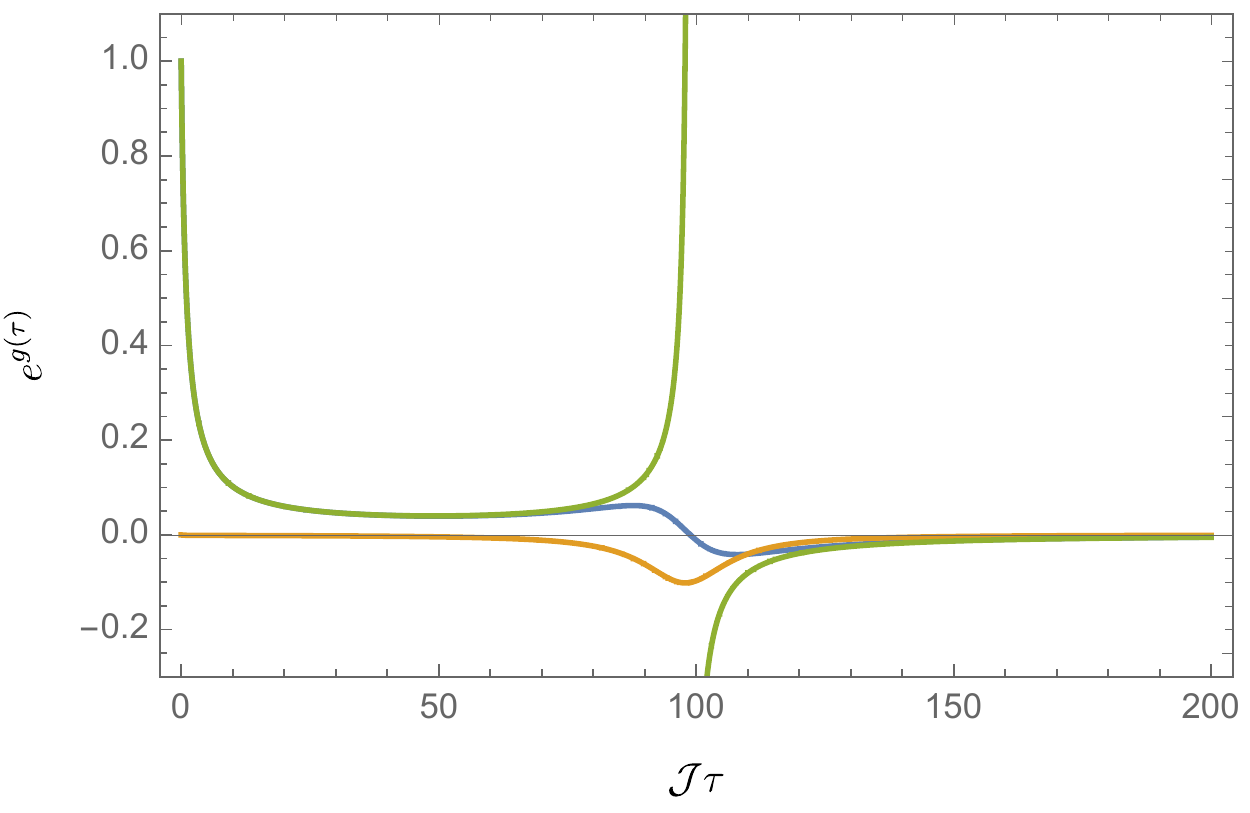} \label{fig_g_1}} 
         \subfigure[]{
                \includegraphics[scale=0.4]{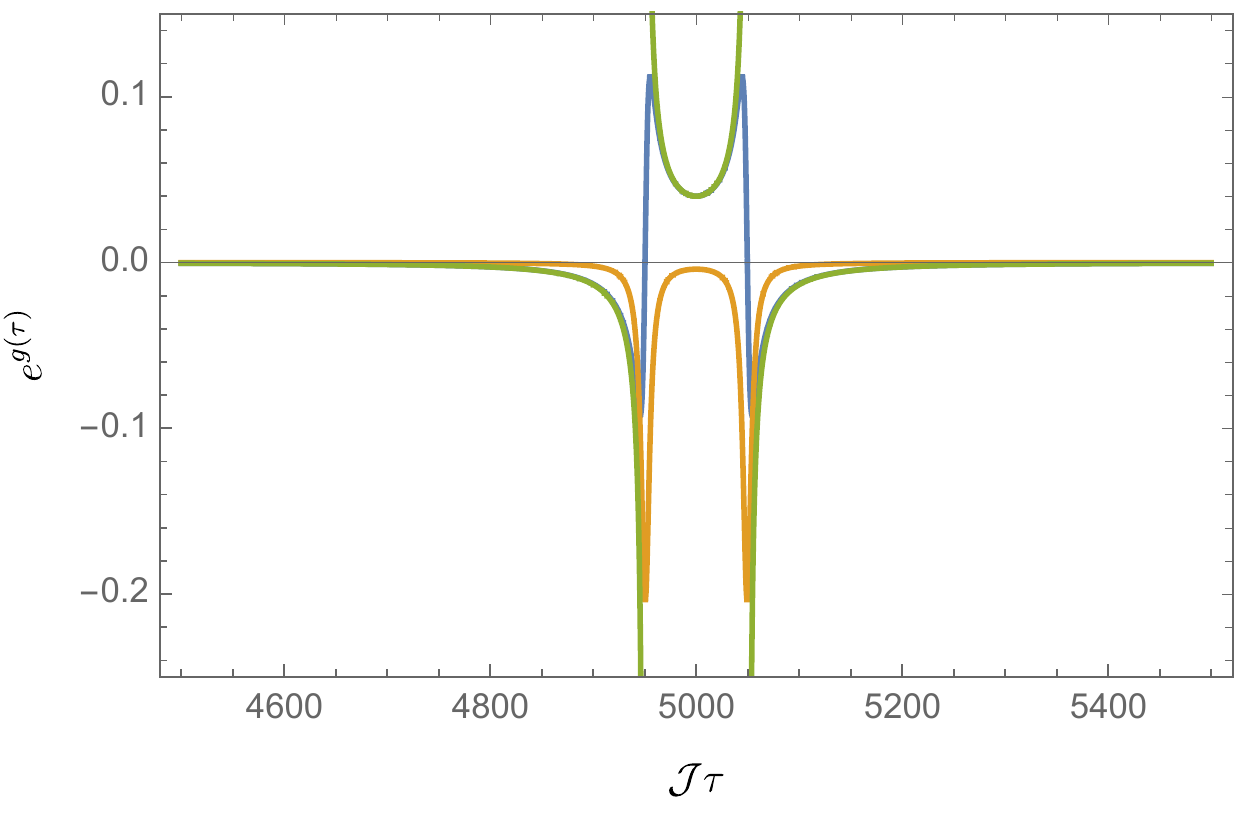} \label{fig_g_2}}  
                     \subfigure[]{
                \includegraphics[scale=0.4]{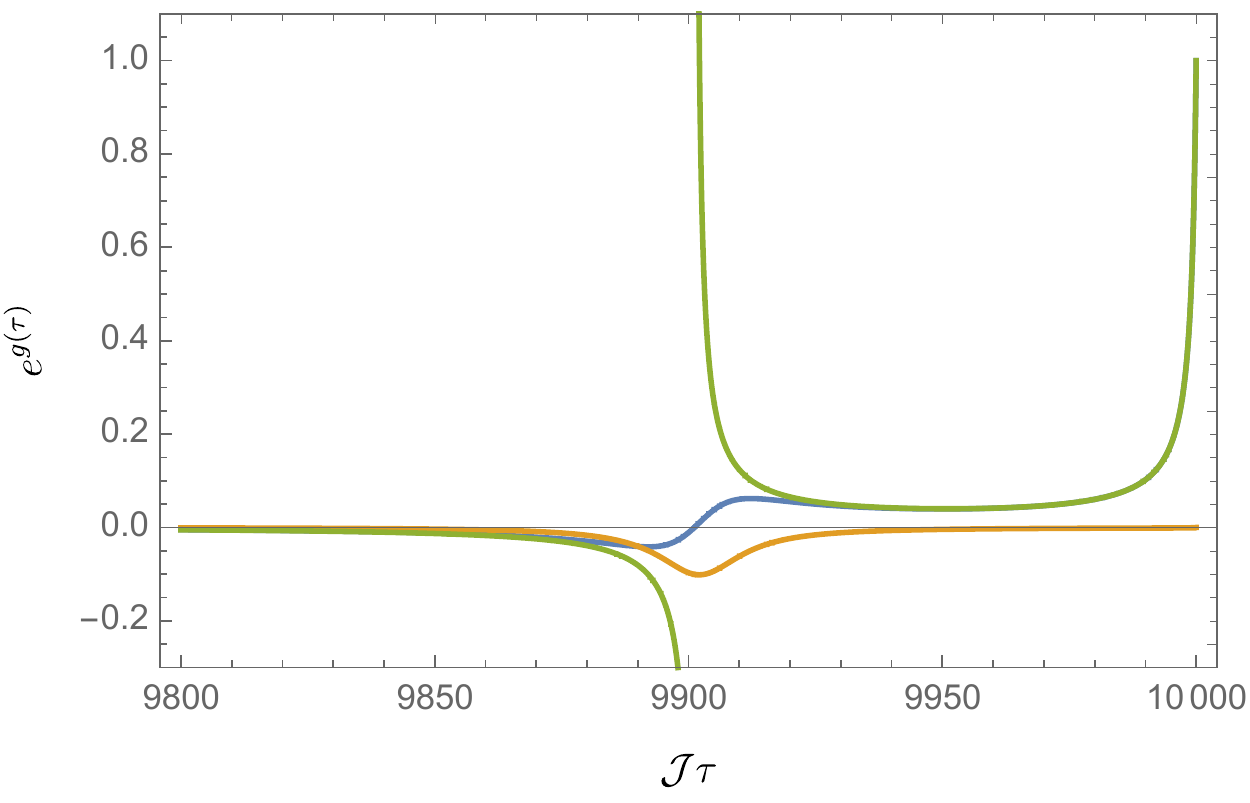} \label{fig_g_3}}
                 \caption{{\footnotesize  The two-point function as a function of time for $\beta \J=10^5$. The green curve corresponds to $s^2=-0.01$, while the blue and yellow curves are the real and imaginary parts when $s^2 = -0.01 +  0.001 i$. We only show the times close to the divergences for the negative coupling and see that the complex coupling gives a smooth two-point function instead.
}}
\label{fig_app_g}
\end{figure}

\begin{figure}[h!]
        \centering
        \subfigure[Free energy for $s^2=-0.01+ 0.001 i$]{
                \includegraphics[scale=0.6]{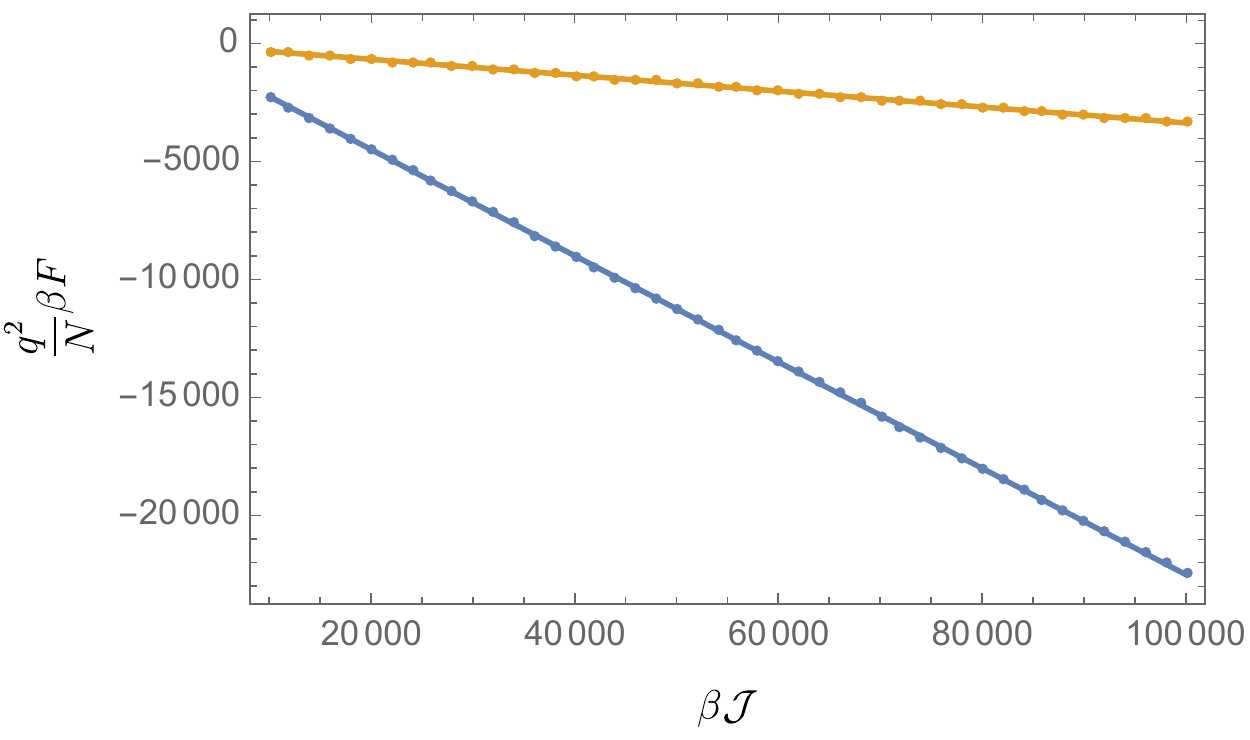} \label{fig_F_n}} \quad\quad
         \subfigure[Specific heat]{
                \includegraphics[scale=0.56]{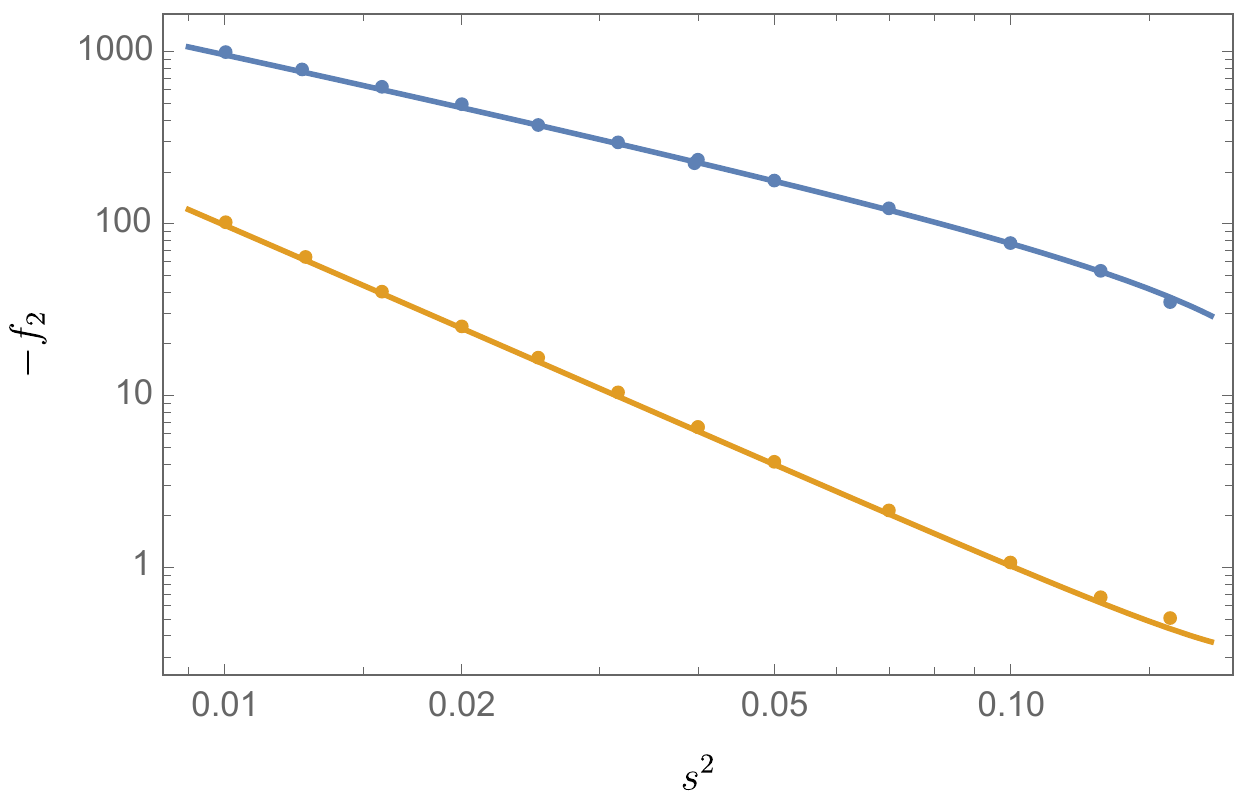} \label{fig_C_n}}
                 \caption{{\footnotesize  In both cases, we plot in blue the real part and in yellow, the imaginary one. (a) The free energy as a function of the inverse temperature for $s^2=-0.01+ 0.001 i $. The dots correspond to numerically integrating (\ref{two_app}) for different temperatures, while the solid line corresponds to the fit by a function of the form (\ref{free_s_negative}). The result is within $1\%$ of the analytic expression (\ref{free_s_negative}), except for the imaginary part of $f_0$. \newline
                 (b) The coefficient $-f_2$ as a function of $s^2$ in a logarithmic scale. The dots correspond to numerically integrating the free energy at each $s^2$ and fitting by a function of the form of (\ref{free_low_temps}). The solid line is the analytic expectation $f_2 = \pi^2 \sqrt{4s^2+1} /s^2$, which gives a negative, linear in temperature specific heat. 
}}
\label{fig_s_negative}
\end{figure}

A way to avoid these divergences is to add a small imaginary part to $s^2$. See figure \ref{fig_app_g}. In that case, the two point function becomes continuous again and we can compute numerically the free energy. The results for $s^2 \to -s^2 + 10^{-3}i $ are shown in figure \ref{fig_s_negative}. The numerical analysis provides substantial evidence that
\begin{equation} \label{free_s_negative}
\beta F = \frac{N}{q^2} \left( -f_0 \beta \J + \pi ^2-\frac{\pi ^2 \sqrt{1+ 4 s^2}}{s^2} \frac{1}{\beta \J} +  \cdots \right) \,,
\end{equation}
which in turns predicts a specific heat
\begin{equation}
C =  \frac{2 \pi ^2 \sqrt{1+ 4 s^2}}{s^2} \frac{N}{q^2 \beta \J} \,,
\end{equation}
that is negative for $-1/4<s^2<0$. Note also that its absolute value is 4 times the one of $s^2>0$. This can be understood from the fact that $\nu$ is close to $2\pi$ in this case.

\bibliographystyle{JHEP}
\bibliography{bibliography}

\end{document}